\documentclass[letterpaper,twocolumn,10pt]{article}
\usepackage{zhanggroup}

\usepackage{hyperref}
\usepackage{xspace}
\usepackage{xurl}
\usepackage{caption}
\usepackage{subcaption}
\usepackage{booktabs}
\usepackage{graphicx}
\usepackage{array}
\usepackage[absolute]{textpos}
\usepackage{float}
\usepackage[dvipsnames]{xcolor}
\hypersetup{
  colorlinks,
  linkcolor={RoyalBlue},
  citecolor={PineGreen},
  urlcolor={Bittersweet}
}

\newcommand{\mypara}[1]{\noindent{\bf {#1}.}\xspace}
\newcommand{\mysubpara}[1]{\noindent{\textit {#1}.}\xspace}
\newcommand{\dataset}{Lexica-Dataset\xspace}
\newcommand{\attack}{PromptStealer\xspace}
\newcommand{\defense}{PromptShield\xspace}

\begin{document}

\begin{textblock}{13}(1.5,1)
\centering
To Appear in the 33rd USENIX Security Symposium, August 2024.
\end{textblock}

\title{Prompt Stealing Attacks Against Text-to-Image Generation Models}

\date{}

\author{
{\rm Xinyue Shen}\ \ \
{\rm Yiting Qu}\ \ \
{\rm Michael Backes}\ \ \
{\rm Yang Zhang}\ \ \
\\
\\
\textit{CISPA Helmholtz Center for Information Security}
}

\maketitle

\begin{abstract}

Text-to-Image generation models have revolutionized the artwork design process and enabled anyone to create high-quality images by entering text descriptions called \emph{prompts}. 
Creating a high-quality prompt that consists of a subject and several modifiers can be time-consuming and costly.
In consequence, a trend of trading high-quality prompts on specialized marketplaces has emerged. 
In this paper, we perform the first study on understanding the threat of a novel attack, namely \emph{prompt stealing attack}, which aims to steal prompts from generated images by text-to-image generation models.
Successful prompt stealing attacks directly violate the intellectual property of prompt engineers and jeopardize the business model of prompt marketplaces.
We first perform a systematic analysis on a dataset collected by ourselves and show that a successful prompt stealing attack should consider a prompt's subject as well as its modifiers.
Based on this observation, we propose a simple yet effective prompt stealing attack, \attack.
It consists of two modules: a subject generator trained to infer the subject and a modifier detector for identifying the modifiers within the generated image.
Experimental results demonstrate that \attack is superior over three baseline methods, both quantitatively and qualitatively.
We also make some initial attempts to defend \attack.
In general, our study uncovers a new attack vector within the ecosystem established by the popular text-to-image generation models.
We hope our results can contribute to understanding and mitigating this emerging threat.\footnote{Our code is available at \url{https://github.com/verazuo/prompt-stealing-attack}.}

\end{abstract}

\section{Introduction}

\begin{figure}[!t]
\centering
\includegraphics[width=.9\linewidth]{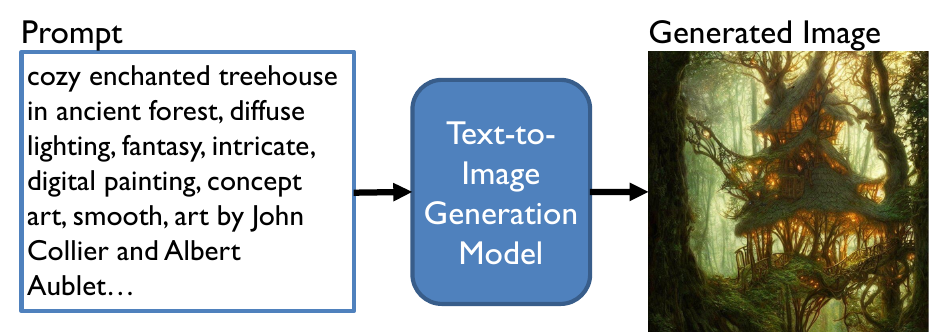}
\caption{An image generated by Stable Diffusion and its corresponding prompt~\cite{RBLEO22}.}
\label{figure:example}
\end{figure}

With the advent of Stable Diffusion~\cite{RBLEO22}, DALL$\cdot$E~2~\cite{RDNCC22}, and Midjourney~\cite{Midjourney}, text-to-image generation models have revolutionized the artwork design process and sparked a surge of artwork creation in mainstream media.
Rather than relying on professional artists, text-to-image generation models empower anyone to produce digital images, such as photorealistic images and commercial drawings, by entering text descriptions called \emph{prompts}.
According to~\cite{O22, LC22, HCDW22}, a high-quality prompt that leads to a high-quality image should consist of a \emph{subject} and several \emph{prompt modifiers}.
The subject is a natural language description of the object or scenarios depicted in the image; the prompt modifiers are keywords or key phrases that are related to specific elements or styles of the image.
\autoref{figure:example} shows an example of a typical prompt and the resulting generated image.
In this example, ``cozy enchanted treehouse in ancient forest'' is the subject and the rest phrases, e.g., ``diffuse lighting,'' ``fantasy,'' etc., are prompt modifiers.

Although it seems that text-to-image generation models simplify the process of artwork design, crafting high-quality prompts is, in fact, complex and iterative.
To create a desired and stable prompt, users need to constantly search for prompt modifiers and check the corresponding resulting images.
Given the limited understanding of the impact of each prompt modifier, this process can be both time-consuming and costly.
As a result, a new job type, namely \emph{prompt engineer}, has emerged for people who specialize in producing high-quality prompts.
Also, high-quality prompts become new and valuable commodities and are traded in specialized marketplaces, such as PromptBase~\cite{PromptBase}, PromptSea~\cite{PromptSea}, and Visualise AI~\cite{VisualiseAI}.
The business model underlying these marketplaces is straightforward: customers browse sample images generated by text-to-image generation models. 
If they like a sample image, they can purchase the corresponding prompt.
Then, customers can freely generate similar images or modify the prompt's subject to generate other images in a similar style.
Prompt marketplaces are gaining popularity.
PromptBase has achieved 10K registered users until November 2022; approximately 45,000 prompts were sold by the top 50 prompt sellers (an estimated \$186,525 in total) over 9 months; new prompt trading platforms, e.g., Prompti AI, Prompt Attack, etc, continue to emerge~\cite{PromptiAI, PromptAttack, PromptBase_1k_users}.

In such a context, a natural research question has emerged: given an image generated by a text-to-image generation model, whether an adversary is able to infer its corresponding prompt?
We name this novel attack as \emph{prompt stealing attack}.
A successful prompt stealing attack directly violates the intellectual property of prompt engineers.
Moreover, it poses a significant threat to the business model of prompt marketplaces.
So far, few tools in the open-source community can be tailored for stealing prompts, such as applying an image captioning model or greedily searching for the modifier combinations with the highest similarity of the image~\cite{ClipInterrogator, MHB21, LLXH22}.
However, it is still unclear whether and to what extent an adversary can effectively steal prompts from generated images.
In this paper, we aim to fill the gap.

\mypara{Our Contributions}
We perform the first large-scale study on understanding the prompt stealing attacks against images generated by text-to-image generation models.
We start by collecting a large-scale dataset from Lexica~\cite{Lexica}, a well-known image gallery with 5M+ prompts and generated images from text-to-image generation models.
Overall, we collect 250,000 pairs of prompts and images.
After preprocessing and de-duplication on prompts, we are left with 61,467 prompt-image pairs (see \autoref{section:data_collection}).
We name the dataset as \emph{\dataset}.
Our quantitative and qualitative analysis on \dataset shows that both a prompt's subject and its modifiers are important factors for the generated image's quality (see \autoref{section:data_analysis}).
This implies a successful prompt stealing attack should consider both subjects and modifiers.

Based on the findings above, we propose a simple yet effective prompt stealing attack, \emph{\attack}.
\attack comprises two modules: a subject generator and a modifier detector. 
Given a target image, the subject generator infers the stolen prompt's subject and the modifier detector predicts modifiers of the stolen prompt simultaneously.
Then, the subject and the prompt modifiers are concatenated as the final stolen prompt.
Experimental results on \dataset show that \attack outperforms the three baseline methods across semantic, modifier, image, and pixel similarities.
For instance, \attack achieves 0.70, 0.43, 0.80, and 0.90 in these metrics, respectively, while the corresponding results for the better baseline CLIP Interrogator~\cite{ClipInterrogator} are 0.52, 0.01, 0.77, and 0.89.
We also qualitatively evaluate \attack and find that the images generated by stolen prompts are similar to the target images.
Our further experiments on real-world prompts traded in marketplaces demonstrate similar results.
Moreover, we show that the performance of \attack can be further boosted by involving the adversary in the loop or providing multiple target images.

Additionally, we make attempts to mitigate prompt stealing attacks by proposing \emph{\defense}.
This method adds an optimized perturbation to an image relying on the technique of adversarial examples such that the adversary cannot infer the image's prompt appropriately.
Experimental results show that \defense works well (see \autoref{section:defense}); however, it requires strong assumptions for the defender, and its performance is reduced by the adaptive attack.

\mypara{Implications}
Our work, for the first time, reveals the threat of prompt stealing in the ecosystem created by the popular text-to-image generation models.
We believe that our findings can serve as a valuable resource for stakeholders to navigate and mitigate this emerging threat. 
Moreover, we hope to raise awareness of academia to work on the safety and security issues of the advanced text-to-image generation models.
We commit to sharing our dataset and code with the research community to facilitate the research in this field.

\mypara{Ethics \& Disclosure}
According to the terms and conditions of Lexica~\cite{Lexica_toc},  images on the website are available under the Creative Commons Noncommercial 4.0 Attribution International License.
We strictly followed Lexica's Terms and Conditions, utilized only the official Lexica API for data retrieval, and disclosed our research to Lexica~\cite{Lexica_toc, Lexica_search_api}.
We also responsibly disclosed our findings to prompt marketplaces such as PromptBase and PromptSea.
PromptBase acknowledged our findings.
They also launched a watermark editor for prompt engineers to defend against prompt stealing attacks.
PromptSea explicitly discusses the risks of prompt stealing attacks in their white paper~\cite{PromptSea_whitepaper}.
Our attack has also been recognized in Microsoft Vulnerability Severity Classification for AI Systems.\footnote{\url{https://www.microsoft.com/en-us/msrc/aibugbar}.}

\begin{figure*}[!t]
\centering
\begin{subfigure}{0.45\linewidth}
\includegraphics[width=\linewidth]{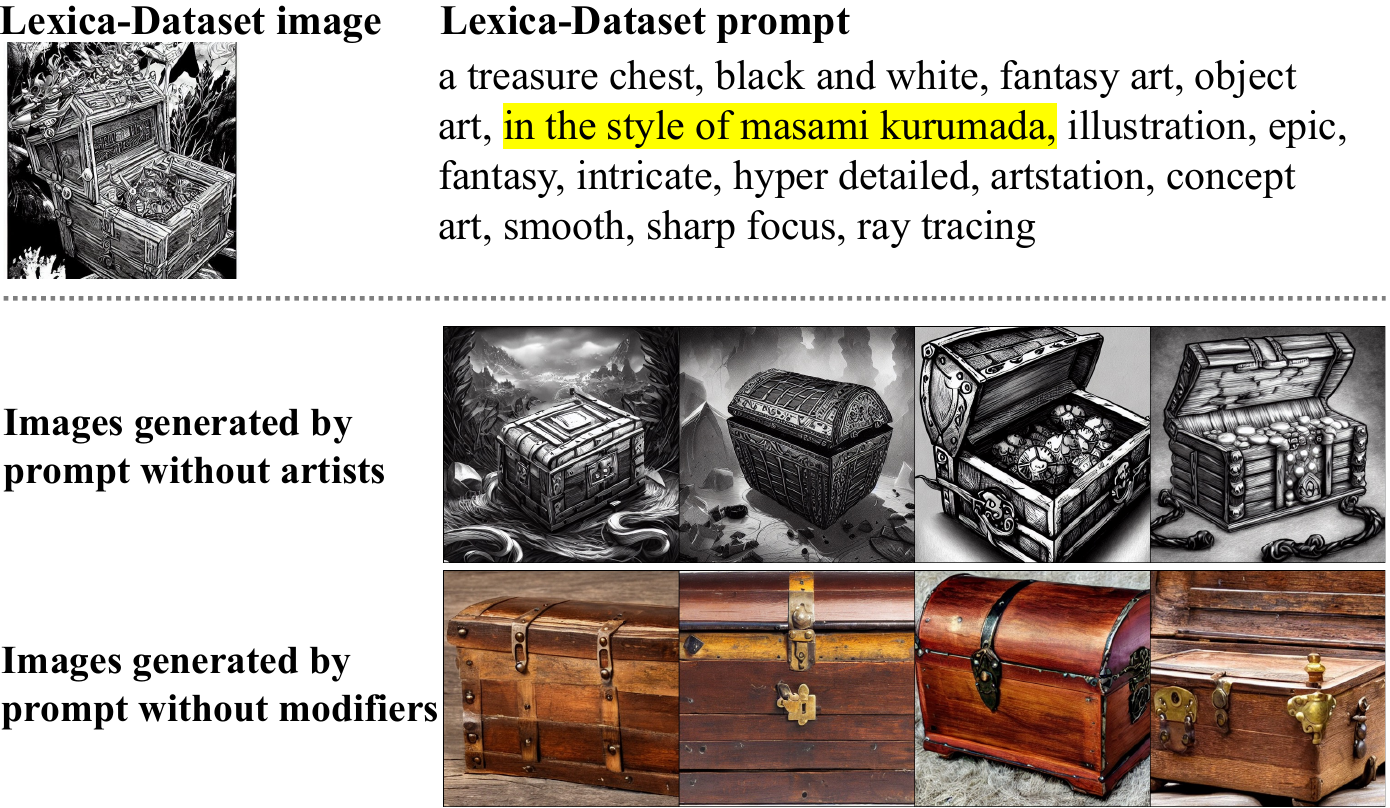}
\caption{}
\label{figure:image_chest}
\end{subfigure}
\begin{subfigure}{0.45\linewidth}
\includegraphics[width=\linewidth]{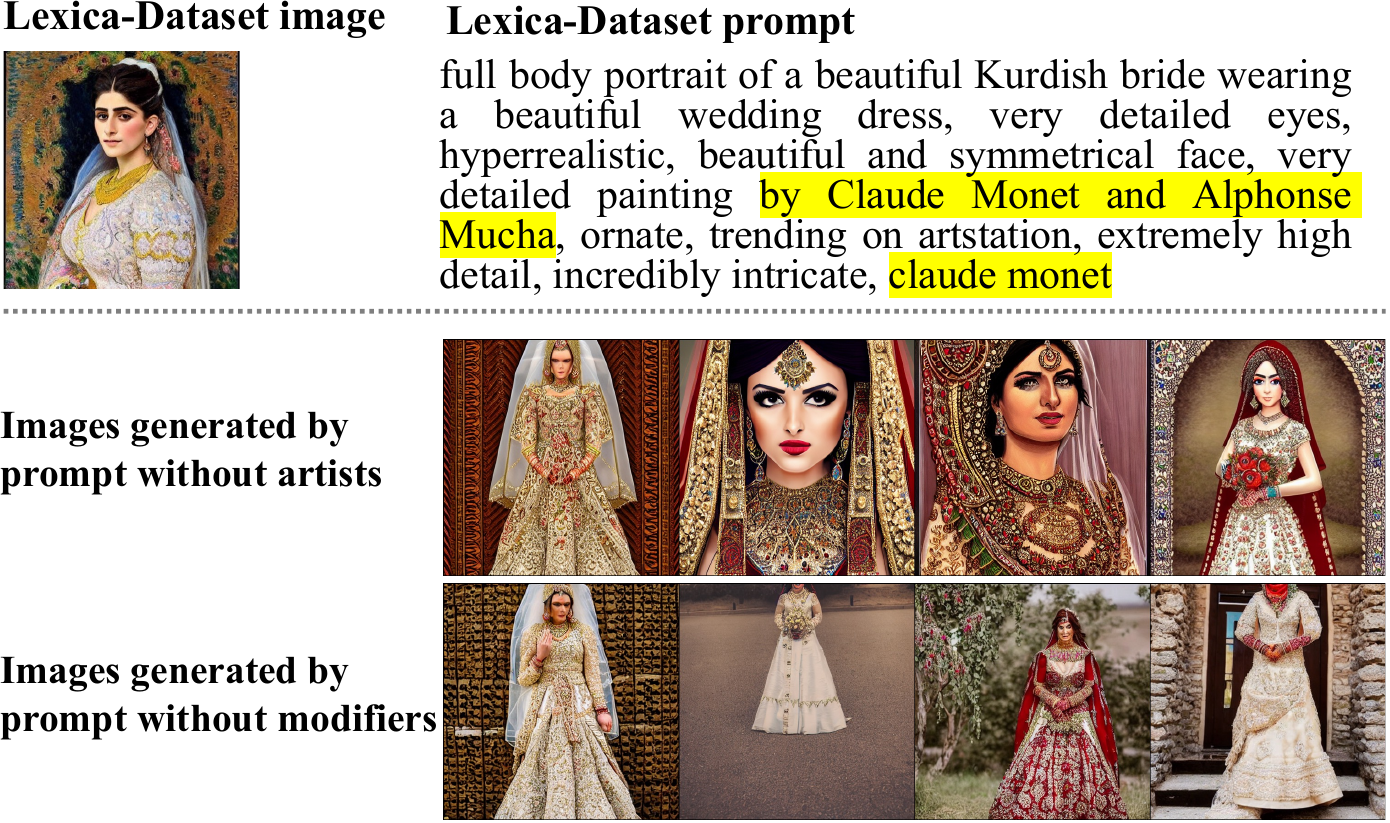}
\caption{}
\label{figure:image_monet}
\end{subfigure}
\caption{Two examples from \dataset, and the corresponding generated images without artist modifiers and without any modifiers.
The yellow highlight represents the artist modifiers.
For the latter two cases, we generate four images with different random seeds to eliminate the potential biases.}
\label{figure:dataset_overview}
\end{figure*}

\section{Background}
\label{section:background}

\subsection{Text-to-Image Generation Models}

Text-to-Image generation models aim to generate high-quality digital images with natural language descriptions, namely prompts.
With the advancement of diffusion models, text-to-image generation has achieved a giant leap and gained significant attention~\cite{RBLEO22, SCSLWDGAMLSHFN22, RDNCC22}.
Images generated by these models have been employed in various scenarios, such as children's books~\cite{children_book}, magazine covers~\cite{magazine_cover}, and fashion imagery~\cite{ai_fashion}.
In this work, we focus on Stable Diffusion, the state-of-the-art and arguably the most popular text-to-image generation model.
Besides, compared to other models like DALL$\cdot$E~2 and Midjounry, Stable Diffusion is open-source and the community that promotes the usage of Stable Diffusion is very active.
Since its release in March 2022, over 110K people have joined the Reddit community to share and discuss images generated by Stable Diffusion~\cite{reddit_StableDiffusion}.

\subsection{The Prompt Marketplace}

While Text-to-Image generation models have gained popularity, producing high-quality images remains a laborious and costly manual process.
A user needs to constantly search for prompt modifiers and check the corresponding resulting images.
One example is the logo of OctoSQL, an open-source CLI project.
The designers use DALL$\cdot$E~2 to create the logo and spend around \$30 interacting with the paid API~\cite{dalle2_octosql_logo}.
As a result, prompt engineers, people who are skilled in producing high-quality prompts, start to sell their prompts in prompt marketplaces.
In these marketplaces, a customer can explore sample images generated by different prompts; if they like a certain image, they can purchase the corresponding prompt through the marketplace.
After getting the prompt, the customer can freely generate similar images or modify the prompt's subject to generate other images in a similar style.

\mypara{Business Volume}
To assess the potential impacts of prompt stealing attacks, we conducted a manual estimation of the business volume of a prompt marketplace, PromptBase. 
We chose PromptBase for two reasons: 
1) PromptBase is the first and most widely recognized prompt marketplace, thus its business volume can represent the marketplaces to some extent; 2) PromptBase discloses the total sales of each prompt engineer. 
In our analysis, we calculate the sales volume of the top 50 prompt engineers over the past 9 months.
We find that they collectively sold approximately 45,000 prompts, resulting in an estimated total revenue of \$186,525.

\mypara{Legal Perspectives of Prompt Trading}
The terms of service and white papers of prompt marketplaces~\cite{PromptBase_terms_of_service, PromptSea_whitepaper} explicitly state that all prompts on their platforms are the intellectual property of the prompt engineers and cannot be obtained through purchase.
Engaging in prompt stealing attacks would, therefore, violate intellectual property rights and jeopardize the business model of the prompt marketplace.

\begin{figure*}[!t]
\centering
\begin{subfigure}{0.3\linewidth}
\includegraphics[width=0.7\linewidth]{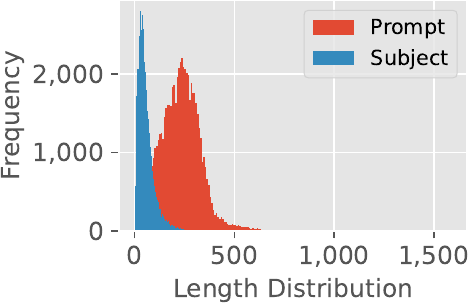}
\caption{Subject/prompt length distribution.}
\label{figure:subject_len}
\end{subfigure}
\begin{subfigure}{0.3\linewidth}
\includegraphics[width=0.7\linewidth]{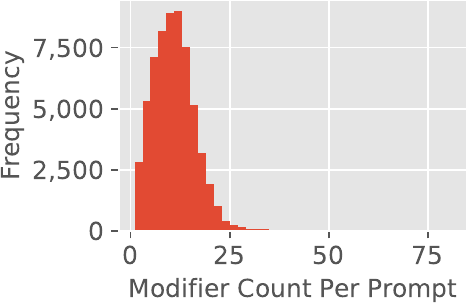}
\caption{Modifier count distribution.}
\label{figure:modifer_count}
\end{subfigure}
\begin{subfigure}{0.3\linewidth}
\includegraphics[width=0.7\linewidth]{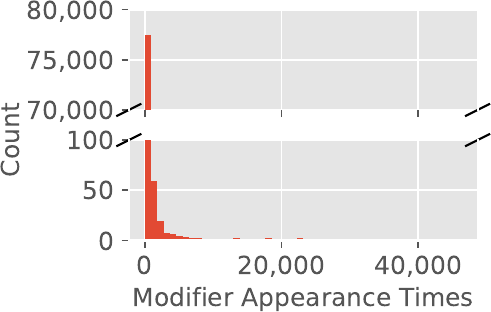}
\caption{Modifier appearance distribution.}
\label{figure:modifier_appearance_times}
\end{subfigure}
\caption{General statistics of prompts in \dataset.}
\end{figure*}

\begin{table*}[!t]
\centering
\caption{Top 10 prompt modifiers and their appearance times for each category.}
\label{table:top10_modifiers}
\tabcolsep 3pt
\scalebox{0.85}{
\begin{tabular}{r|lr|p{.2\linewidth}r|p{.2\linewidth}r|lr|lr}
\toprule
\textbf{No.} & \textbf{Trending} & \multicolumn{1}{r|}{\textbf{\#}} & \textbf{Artist} & \multicolumn{1}{r|}{\textbf{\#}} & \textbf{Medium} & \multicolumn{1}{r|}{\textbf{\#}} & \textbf{Movement} & \multicolumn{1}{r|}{\textbf{\#}} & \textbf{Flavor} & \multicolumn{1}{r}{\textbf{\#}}\\
\midrule
1 & artstation & 46,357 & greg rutkowski & 11,552 & concept art & 24,095 & fantasy art & 1,567 & highly detailed & 26,818\\ \hline
2 & cgsociety & 3,822 & artgerm & 7,273 & digital art & 7,483 & art nouveau & 1,132 & 8k & 22,987\\ \hline
3 & deviantart & 2,309 & artgerm and greg rutkowski and alphonse mucha & 6,237 & vector art & 127 & hyperrealism & 1,041 & sharp focus & 22,774\\ \hline
4 & pixiv & 1,018 & wlop & 5,989 & an ultrafine detailed painting & 83 & photorealism & 924 & digital painting & 18,211\\ \hline
5 & pinterest & 646 & ilya kuvshinov & 3,918 & a detailed matte painting & 63 & street art & 447 & intricate & 18,002\\ \hline
6 & behance & 321 & alphonse mucha & 3,716 & poster art & 58 & surrealism & 388 & illustration & 17,473\\ \hline
7 & instagram & 127 & rossdraws & 2,802 & an oil painting & 51 & romanticism & 384 & octane render & 14,394\\ \hline
8 & zbrush central & 57 & craig mullins & 2,341 & pixel art & 42 & art deco & 377 & smooth & 13,906\\ \hline
9 & cg society & 52 & james jean & 2,133 & a detailed painting & 40 & realism & 292 & elegant & 13,081\\ \hline
10 & polycount & 51 & rhads & 1,976 & a character portrait & 39 & rococo & 290 & 4k & 10,325\\   
\bottomrule
\end{tabular}
}
\end{table*}

\section{Preliminary Analysis}
\label{section:measurement}

\subsection{Data Collection}
\label{section:data_collection}

To the best of our knowledge, no large-scale prompt-image datasets are available at the time of our study.
In order to assess prompt stealing attacks, we collect a dataset by ourselves via Lexica~\cite{Lexica}.
We opt for Lexica for two reasons.
First, Lexica is a well-known image gallery of Stable Diffusion; it contains over 5M prompt-image pairs extracted from Stable Diffusion's discord server~\cite{discord_StableDiffusion}.
Lexica covers the art creations of many Stable Diffusion's active users, such as artists and prompt engineers.
This allows us to better simulate the real-world attack scenario.
Second, Lexica provides a query API~\cite{Lexica_search_api} that given a specific prompt, returns the 50 most similar prompts and the corresponding images.
Note that similar prompts are likely from the same user during their prompt engineering process.
Therefore, the data from Lexica can closely reflect people's real-world usage of Stable Diffusion.

As Lexica only provides a query API and does not release its public dataset, we regard an open-source prompt dataset on Hugging Face as a starting point for our data collection.\footnote{\url{https://huggingface.co/datasets/Gustavosta/Stable-Diffusion-Prompts}.}
This dataset contains 80K prompts crawled from Lexica, which are used to train a prompt generator.
However, since this dataset does not contain any images and the image download link, we cannot directly use it.
To address this, we randomly sample 5,000 prompts from the dataset and then query these prompts to Lexica's query API.
As mentioned above, the API returns 50 results for each query, including images, prompts, grid, etc.
In the end, we crawl 250,000 prompts and the corresponding generated images.
We specifically exclude images generated by models other than Stable Diffusion, such as Aperture, a model designed to generate photorealistic images and can only be accessed on the Lexica website.
We filter out the prompts and images that could not be parsed correctly.
Also, images belonging to the grid type (each of those images contains several generated images stitched together) are neglected as well.
Finally, after the de-duplication of prompts, we get 61,467 prompt-image pairs.
We name the dataset as~\emph{\dataset}.

As mentioned before, each prompt consists of a subject and several prompt modifiers.
Thus, we further decompose prompts in \dataset into subjects and prompt modifiers.
Concretely, we split each prompt by commas.
The first part is regarded as the subject according to~\cite{O22,LC22}, while the rest are treated as prompt modifiers.
We standardize the format of some modifiers, e.g., ``3 d'' to ``3d.''
We also remove the style-evocation words, including ``in the style of,'' ``inspired by,'' ``trending on,'' etc.
In the end, we obtain 77,616 prompt modifiers.
\autoref{figure:dataset_overview} depicts some samples from the dataset.

\subsection{Data Analysis}
\label{section:data_analysis}

\mypara{Subjects and Prompt Modifiers}
\autoref{figure:subject_len} depicts the length distribution of subjects and prompts in \dataset.
On average, the length of a subject is 56 characters, while the length of a prompt is 237 characters.
This means a subject takes only 23.63\% of a prompt, and the rest is assembled with prompt modifiers.
\autoref{figure:modifer_count} further depicts the distribution of modifier count per prompt.
On average, each prompt contains 11 modifiers.
This might suggest prompt modifiers play an essential role in the image generation process.
To further investigate this, \autoref{figure:dataset_overview} also shows some generated images without prompt modifiers.
As we can see, modifiers indeed largely influence image quality in terms of style and details.
Take the treasure chest as an example (see \autoref{figure:image_chest}), if the modifiers are not considered, Stable Diffusion generates a plain treasure chest.
Meanwhile, by considering various modifiers, the original image depicts a more stylized treasure chest with many fine-grained details.
These results imply that a successful prompt stealing attack should not only consider the target prompt's subject but also put efforts into recovering the modifiers.
In other words, using an image captioning model to get the caption (subject) of a target image is insufficient for prompt stealing.
Experiment results in \autoref{section:attack} further confirm this.
On the other hand, we emphasize that the subject also plays an essential role in a prompt as it describes the main content of the image.

For the modifiers, we further show the distribution of their appearance times in \autoref{figure:modifier_appearance_times}.
We can observe a Pareto distribution.
Specifically, \dataset contains 61,467 prompts and 77,616 modifiers.
Among them, only 7,672 (9.88\%) modifiers are used more than ten times, and 1,109 (1.43\%) modifiers are used more than 100 times.
From the adversary's perspective, this Pareto distribution eases the requirements for prompt stealing attacks, as a relatively small modifier set can already cover most modifiers used in prompts.

\mypara{Modifiers in Category-Level}
We further attribute prompt modifiers into different categories.
The open-source prompt engineering tool CLIP Interrogator~\cite{ClipInterrogator} defines five categories for modifiers, i.e., trending, artist, medium, movement, and flavor, and offers a relatively complete modifier list for each category.
In our analysis, we adopt the same set of categories and assign all modifiers in \dataset to them.
Concretely, a modifier in \dataset is considered as part of a category if it belongs to the corresponding modifier list from CLIP Interrogator.
\autoref{table:top10_modifiers} shows the top 10 prompt modifiers with respect to their appearance times in each category of \dataset.
We find that users are less likely to use movement modifiers.
The highest one is ``fantasy art'' with appearance times being only 1,567.
This is probably because the usage of movement modifiers requires background knowledge of art.
Conversely, the top three modifiers in the flavor category are all well-known image quality boosters~\cite{O22}, such as ``highly detailed'' and ``8k.''
On average, 14.32\%, 6.11\%, 3.25\%, 0.98\%, and 75.33\% of the modifiers in a prompt belong to the artist, trending, medium, movement, and flavor categories, respectively.
Excluding the flavor category which contains various kinds of modifiers, the artist category occupies the largest proportion of prompts compared to the rest.
This reveals that users are more likely to use artist modifiers to lead the generated images to certain visual styles.
\autoref{figure:dataset_overview} further shows some examples of generated images without artist modifiers.
In \autoref{figure:image_monet}, the artist modifiers ``Claude Monet'' and ``Alphonse Mucha'' indeed heavily influence the final outlook and fine-grained details of the original image, compared to the images without any artist modifiers.
A similar observation can be made in the example depicted in \autoref{figure:image_chest}.

\begin{figure}[!t]
\centering
\includegraphics[width=0.6\linewidth]{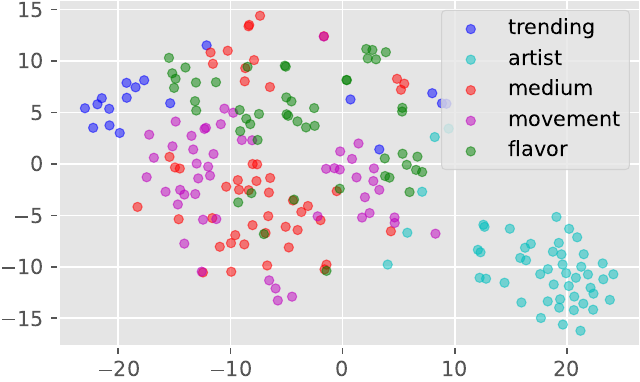}
\caption{T-SNE visualization of modifier embeddings with respect to different categories.}
\label{figure:modifier_tsne}
\end{figure}

Next, we investigate the semantic relations among modifiers of different categories.
Diffusion-based text-to-image generation models like Stable Diffusion rely on CLIP~\cite{RKHRGASAMCKS21} to obtain text embeddings of prompts.
We thereby utilize CLIP's text encoder to obtain modifiers' embeddings and visualize them via a T-SNE~\cite{MH08} plot (see \autoref{figure:modifier_tsne}).
For each category, we consider the top 20 modifiers with respect to their appearance times.
We find that the modifiers in the artist category are tightly grouped and distant from others.
Meanwhile, the other four categories are relatively mixed.
This again shows the importance of the artist modifiers in the image generation process.
Meanwhile, it is expected that other categories are close to each other.
For instance, movement modifiers are often accompanied by medium modifiers.
This has also been uncovered in a previous qualitative study~\cite{LC22} and recommended by DALL$\cdot$E~2 prompt book~\cite{Dalle2PromptBook}.

\mypara{Takeaways}
To summarize, we have established a prompt-image dataset \dataset that can closely reflect people’s real-world usage of Stable Diffusion.
Our data analysis reveals that besides its subject, a prompt's modifiers also play an essential role in determining its generated image's outlook.
This suggests a successful prompt stealing attack should consider both the subject and modifiers of a prompt.
Moreover, we attribute prompt modifiers into five different categories and discover that the artist modifiers heavily influence generated images' styles and details.

\section{Prompt Stealing Attack}
\label{section:attack}

\begin{figure}[!t]
\centering
\includegraphics[width=\linewidth]{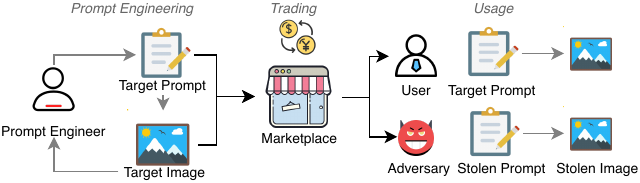}
\caption{The scenario of prompt trading and stealing.}
\label{figure:attack_overview}
\end{figure}

\subsection{Threat Model}
\label{section:threatmodel}

\mypara{Scenario}
Our attack is designed for the scenario shown in \autoref{figure:attack_overview}.
First, a prompt engineer performs an iterative exploring process to find an ideal prompt that can lead to a high-quality image using a text-to-image generation model.
This prompt is referred to as the \emph{target prompt}.
Then, the prompt engineer can sell the prompt via marketplaces like PromptBase~\cite{PromptBase}, by providing an example image generated by the target prompt, namely \emph{target image}.
If a regular user is interested in the style of the target image, the user can get the target prompt by trading with the prompt engineer through the marketplace.
After getting the target prompt, one of the usage scenarios for the user is to modify the prompt's subject to generate other images with a similar style (see \autoref{section:background}).

Conversely, an adversary aims to steal the target prompt from a target image.
We refer to the prompt stolen by the adversary as \emph{stolen prompt}.
Moreover, the adversary can feed the stolen prompt to the text-to-image generation model again, and the generated image here is referred to as \emph{stolen image}.

\mypara{Adversary's Goal}
Given a target image generated by a text-to-image generation model, the goal of the adversary is to steal the target prompt that leads to the target image.
The stolen prompt should ideally satisfy four quantitative goals.
\begin{itemize}
\item \mypara{Semantic Similarity} 
The stolen prompt should share high semantic similarity with the target prompt, normally measured in the text embedding space.
This is important because text-to-image generation models essentially condition on the text embeddings during the image generation.
Here, the semantic similarity considers both subjects and prompt modifiers.
\item \mypara{Modifier Similarity} 
The stolen prompt should contain as many modifiers of the target prompt as possible.
As the modifiers take an essential role in guaranteeing the quality of the generated image (see \autoref{section:measurement}), higher modifier similarity results in better attack performance.
\item \mypara{Image / Pixel Similarity} 
By feeding the stolen prompt to the text-to-image generation model again, the adversary expects the model to generate a similar stolen image to the target image.
Here, the similarity can be regarded as both image semantic and pixel-level similarity.
\end{itemize}
Besides, considering the usage of the target prompt mentioned in \autoref{section:background}, the stolen prompt should also lead to images depicting different subjects in the style of the target image.
Therefore, we also qualitatively assess \attack from this angle. 

\mypara{Adversary's Capability}
We assume the adversary's capabilities in a real-world setting.
First, the adversary can collect public prompts and the corresponding images via online services like Lexica.
Second, the adversary has black-box access to the target text-to-image generation model.

\subsection{\attack}

\begin{figure}[!t]
\centering
\includegraphics[width=\linewidth]{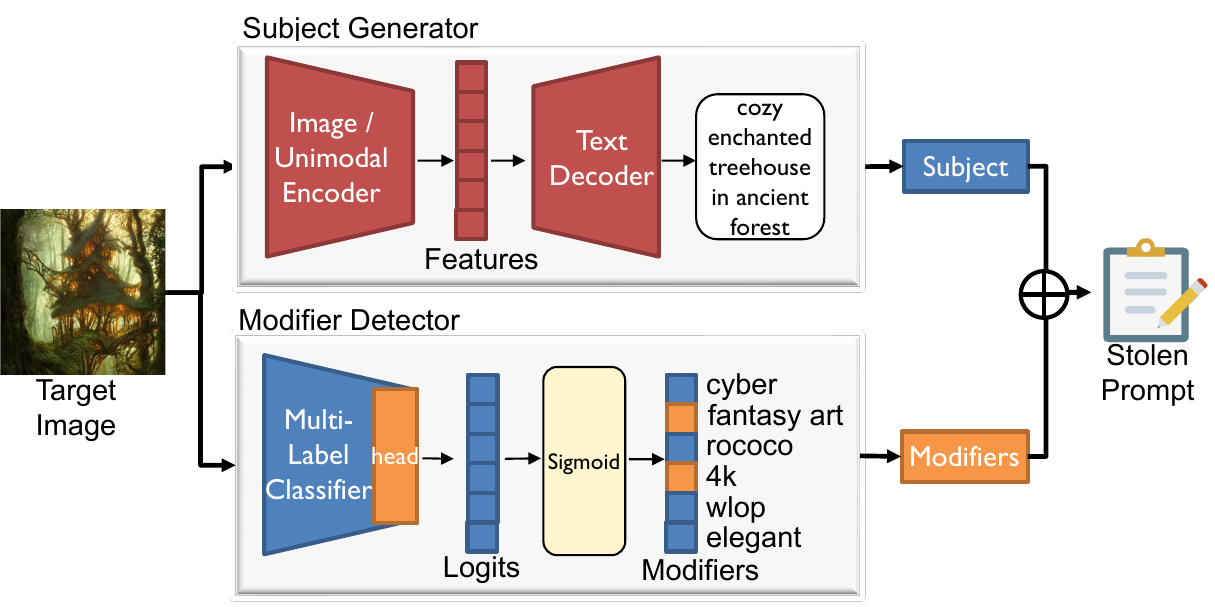}
\caption{Overview of \attack.}
\label{figure:attack_arch}
\end{figure}

In this work, we introduce \attack, a simple yet effective prompt stealing attack.
The design principle of \attack is based on our observations in \autoref{section:measurement}, i.e., a successful prompt stealing attack should focus on both the subject and modifiers of a target prompt. 
Specifically, \attack consists of two modules: a subject generator and a modifier detector.
\autoref{figure:attack_arch} shows the overview of \attack.

\mypara{Subject Generator}
Given the target image, the subject generator aims to generate the subject of the stolen prompt.
In this work, we adopt the image captioning framework for this purpose, which typically consists of an image/unimodal encoder and a text decoder~\cite{LLXH22, WYHLLGLLW22}.
Concretely, the encoder first converts the target image to a fixed-length feature representation.
This representation is then used in a text decoder to generate a caption/subject for the target image.
Note that existing image captioning models are not directly suitable for subject generation for two main reasons. 
Firstly, they are trained on normalized data and are prone to use generic terms, such as referring to a specific celebrity as ``a woman'' or labeling a ``terrier dog'' simply as ``a dog.''
In contrast, the subject of a high-quality prompt demands more specific and detailed information, such as race and landmarks.
Secondly, image captioning models are trained to describe real-world pictures; however, the target images often showcase various art styles.
To bridge this gap, we fine-tune the encoder as well as the decoder on the subjects and corresponding images in \dataset.
We evaluate two representative image captioning frameworks, i.e., BLIP and GIT, in our experiments (see \autoref{section:attack_ablation}).

\mypara{Modifier Detector}
Modifiers are keywords or key phrases related to specific elements or styles of the target image.
A high-quality prompt typically includes numerous and varied modifiers that complexly interact with each other to dictate the style of the generated image.
The goal of the modifier detector is thus to detect modifiers in a given target image.
A similar task in the CV domain is object detection, which leverages a multi-label classifier to predict all objects in one given image simultaneously.
Existing multi-label classifiers commonly consist of a backbone model and a multi-label classification head.
The backbone model outputs an image representation, and the classification head transforms the representation into prediction logits.
However, existing multi-label classifiers cannot be directly applied in the scenario of modifier detection.
This is because they are primarily designed to output labels like ``cat,'' ``dog,'' and ``airplanes,'' which are not modifiers such as artist names or painting styles.
To adapt multi-label classifiers to our scenario, we reset the multi-label head layer of the multi-label classifier with \dataset's modifier set and then fine-tune the whole model on the modifier sets and corresponding images in \dataset.
In addition, we apply a Sigmoid activation layer at the end of the multi-label classifiers to normalize the outputs to prediction posteriors.
The labels/modifiers whose posteriors are higher than a predefined threshold are regarded as the stolen prompt's modifiers.

After applying the subject generator and the modifier detector to the target image, we concatenate the obtained subject and the prompt modifiers together as the final stolen prompt.

\subsection{Experimental Settings}
\label{section:exp_settings}

\mypara{Text-to-Image Generation Model}
We focus on Stable Diffusion as it is one of the most prevalent text-to-image generation models in the field.
Besides, the open-source nature and the active community support (e.g., Lexica) of this model enable us to perform a large-scale evaluation.
Most of our experiments are directly performed on \dataset.
In certain cases, we also need to use Stable Diffusion to generate images for evaluation.
To this end, we adopt the official Stable Diffusion v1.4 model.\footnote{\url{https://huggingface.co/CompVis/stable-diffusion-v1-4}.}
For each image generation process, we sample 50 steps with default settings~\cite{WLZTKC18}.
The size of the generated image is 512$\times$512.
We also consider other representative text-to-image generation models, like DALL$\cdot$E~2 and Midjounery.
However, since they are not open-source and only provide non-free APIs, collecting large-scale datasets from them is financially infeasible for us.
As an alternative, we directly apply \attack trained on \dataset (based on Stable Diffusion) to them as a case study to test our approach's generalizability (see \autoref{section:case_study}).

\mypara{\attack}
\attack consists of two modules: a subject generator and a modifier detector.
For the subject generator, we evaluate two representative image captioning frameworks, i.e., BLIP~\cite{LLXH22} and GIT~\cite{WYHLLGLLW22}.
Concretely, we first adopt the BLIP and GIT model pre-trained on MS-COCO dataset and then follow their default settings to fine-tune them on the subjects and corresponding images of \dataset.
Regarding the modifier detector, we assess two multi-label frameworks, ML-Decoder~\cite{RSBBN21} and Query2Label~\cite{LZYSZ21}.
Note, we have a total of 77,616 prompt modifiers in \dataset, which can all be treated as labels for the multi-label classifier.
However, many modifiers only appear a few times (see \autoref{figure:modifier_appearance_times}).
Therefore, in our main experiments, we only consider modifiers that appear more than 10 times (7,672 modifiers) as labels.
We choose 0.6 as the threshold to decide whether a label/modifier is in the modifier set of the stolen prompt.
The impacts of different label numbers and thresholds are investigated in \autoref{section:attack_ablation}.
\attack is trained on 80\% of the samples in \dataset, and the rest samples are used for testing.

\mypara{Baseline Methods}
We consider three baseline attack methods in our experiments.
The first baseline is a pre-trained image captioning model.
Given an image, the model produces a caption for the image, which is directly regarded as the stolen prompt.
Here, we regard BLIP pre-trained on MS-COCO dataset as the first baseline.
We further fine-tune the first baseline model on \dataset as our second baseline.
Besides, we also include an open-source prompt engineering tool, CLIP Interrogator~\cite{ClipInterrogator} in our evaluation.
This method iteratively calculates the similarity between the combinations of modifiers and the target image.
Once the similarity stops rising, it regards the current combination as the stolen prompt.
Before comparison, we evaluate the performance of CLIP Interrogator with two modifier sets, i.e., the same modifier set as \attack (7,637 modifiers) and the complete modifier set of \dataset (77,616 modifiers).
As shown in \autoref{section:clip_inter_modifiers}, CLIP Interrogator with larger prompt modifiers demonstrates slightly better performance, therefore we utilize CLIP Interrogator with the complete modifier set as the third baseline method.

\mypara{Evaluation Metric}
As discussed in \autoref{section:threatmodel}, the adversary has four quantitative goals: semantic similarity, modifier similarity, image similarity, and pixel similarity.
Thus, we adopt four quantitative metrics for these goals, respectively.

\begin{itemize}
\item \mypara{Semantic Similarity}
The semantic similarity is the cosine similarity between the embeddings of the target prompt and the stolen prompt.
We use CLIP's text encoder to get the embeddings.
\item \mypara{Modifier Similarity} 
The modifier similarity is the Jaccard similarity between the modifiers of the target prompt and those of the stolen prompt.
\item \mypara{Image Similarity} 
The image similarity is the cosine similarity between the embeddings of the target and stolen images, which is a widely adopted metric to measure similarity between images, both at the semantic level and the visual/pixel level~\cite{SCWZHZ23}.
Following previous works~\cite{SCWZHZ23,BHE23}, we rely on CLIP's image encoder to obtain an image's embedding.
For each stolen prompt, we generate four stolen images from Stable Diffusion (by varying the random seed) to eliminate the potential biases.
We then calculate the similarity between each stolen image and the target image and, in the end, average the results.
\item \mypara{Pixel Similarity} 
The pixel similarity is a traditional metric to quantify the pixel differences between the target image and the stolen, which are commonly calculated by the complement of the mean squared error (MSE)~\cite{WBSS04}.
\end{itemize}
We acknowledge that any metric has limitations.
To perform a comprehensive assessment, we also provide qualitative evaluations to assess the similarity perceived by the end-users.

\begin{itemize}
\item \mypara{Human-Rated Similarity} The human-rated similarity refers to the perceived similarity between target and stolen images by end-users.
Specifically, for each target image and its corresponding stolen images, two labelers are assigned to label it using a 5-level Like-scaler, ranging from ``not similar at all'' to ``very similar.''
The detailed criteria for each level are stated in \autoref{table:human_rated_rules} in the Appendix.
We randomly sample 10 target images in our test set and report the mean value.
\end{itemize}

\begin{figure*}[!t]
\centering
\begin{subfigure}{0.45\linewidth}
\includegraphics[width=\linewidth]{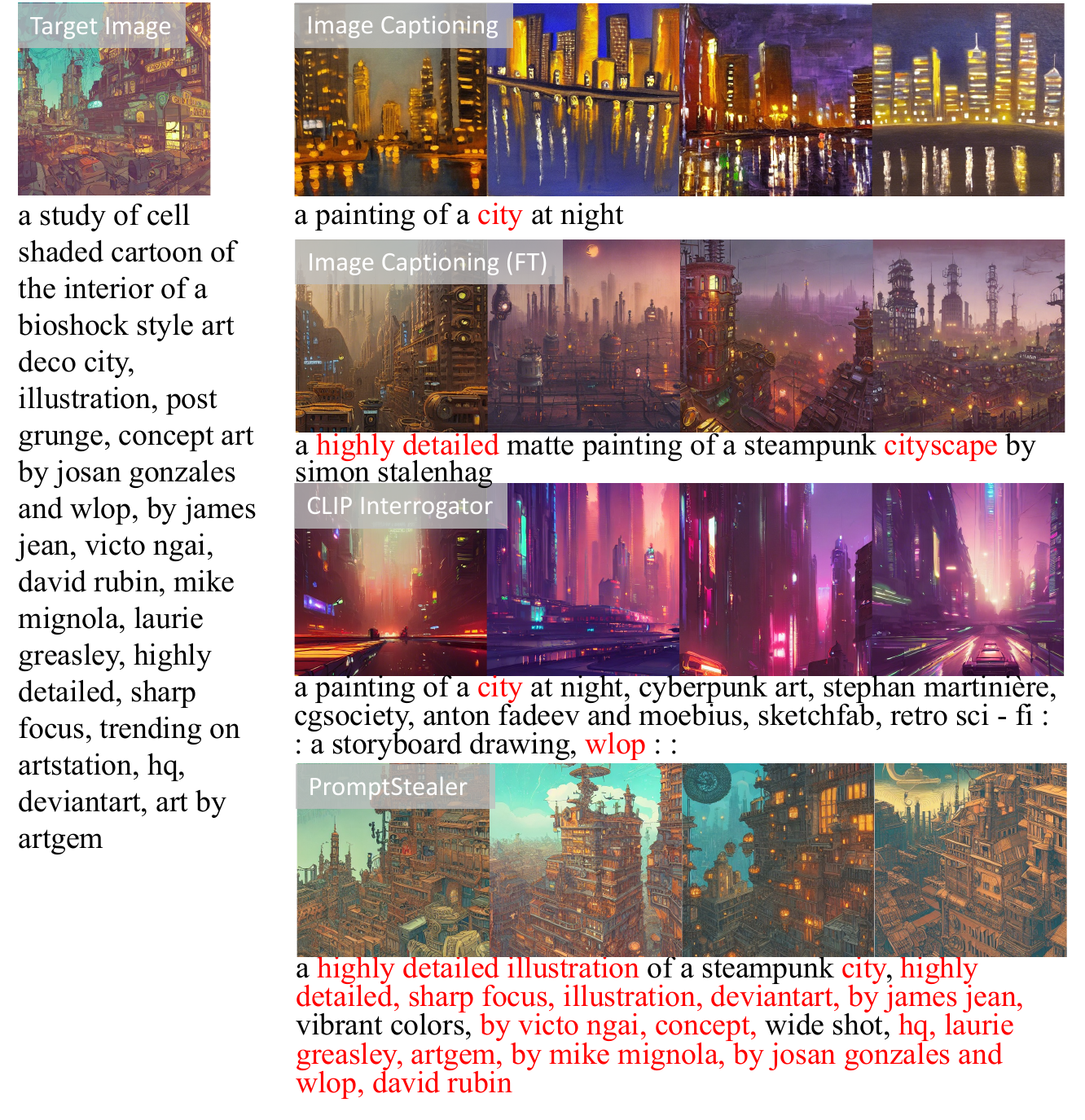}
\caption{}
\label{figure:attack_example_a}
\end{subfigure}
\begin{subfigure}{0.45\linewidth}
\includegraphics[width=\linewidth]{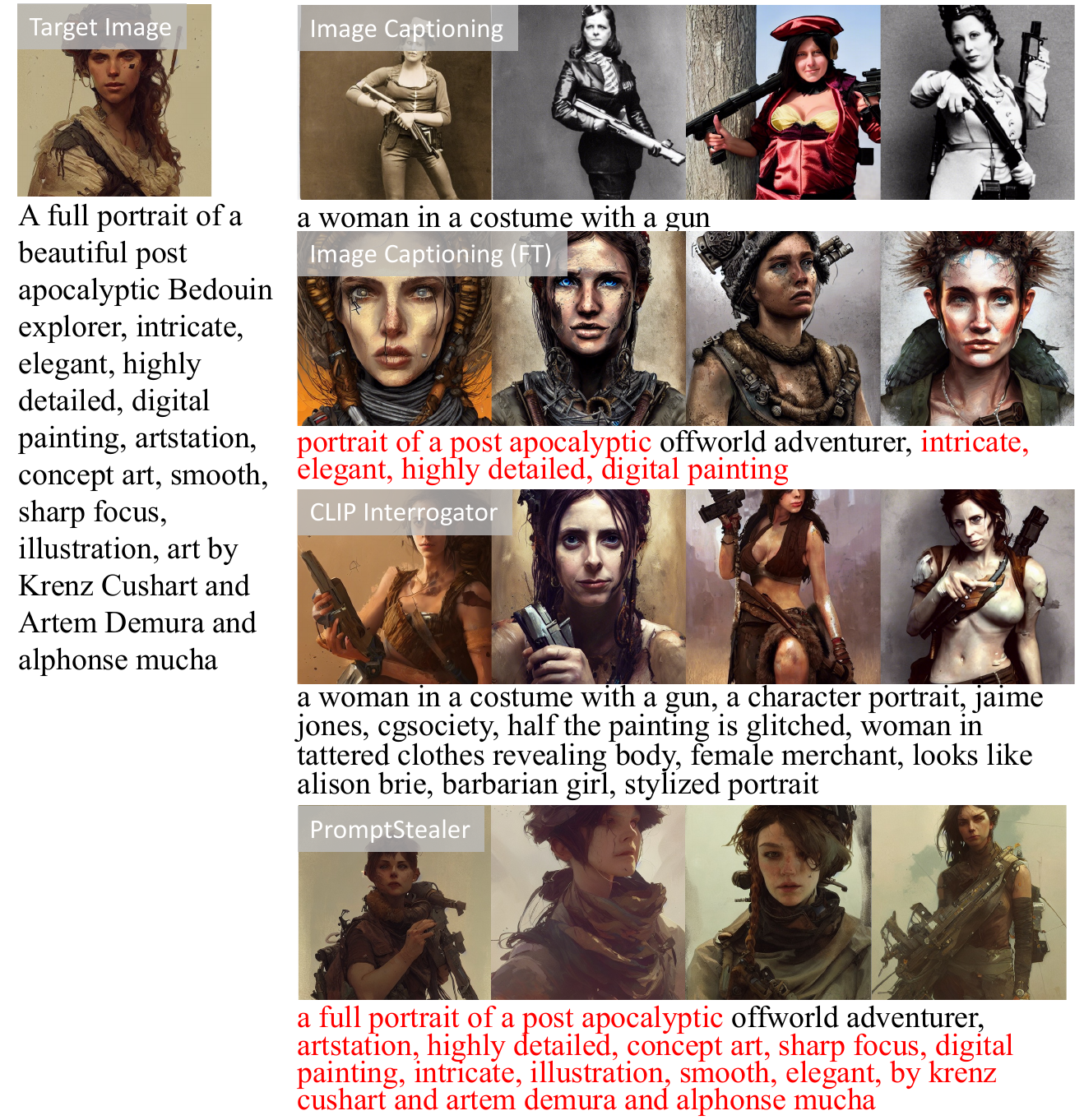}
\caption{}
\label{figure:attack_example_b}
\end{subfigure}
\caption{Two attack examples of \attack and the three baselines.
The text below each image is the target/stolen prompt.}
\label{figure:attack_exmaples}
\end{figure*}

\subsection{Quantitative Evaluation}

\autoref{table:main_results_lexica} shows the performance of \attack together with the three baselines.
We first observe that the image captioning model itself is not sufficient to achieve a successful prompt stealing attack. 
Even fine-tuned on the \dataset, it only achieves 0.45, 0.14, 0.74, and 0.89 in semantic, modifier, image and pixel similarities.
To compare, \attack achieves 0.70, 0.43, 0.80, and 0.90 in these metrics, respectively, while the corresponding results for the better baseline CLIP Interrogator are 0.52, 0.01, 0.77, and 0.89.
We also notice that the modifier similarity of \attack is not as high as other metrics.
This is primarily because the modifier similarity is based on exact word matching, which is typically not high in the NLP domain~\cite{H13}.
For example, modifiers like ``digital art'' and ``digital painting'' are not considered a match despite their high semantic similarity.
To address this, we offer semantic similarity as an additional metric.
We further observe that the pixel similarity is not very distinguishable in this scenario.
For instance, \attack performs slightly better than CLIP Interrogator (0.90 vs. 0.89 in pixel similarity).
This might be due to the pixel similarity metric itself.
Since it is sensitive to changes in pixel values, it does not necessarily consider the human perception of similarity.
We have evaluated other pixel-level metrics such as SSIM~\cite{WBSS04} and pHash~\cite{S17}.
The performance is similar.
Therefore, we introduce human-rated similarity in the qualitative evaluation.

\begin{table}[!t]
\centering
\caption{The performance of \attack and the three baselines on \dataset. 
ImgCap refers to the image captioning method; ImgCap (FT) represents the image captioning method fine-tuned on \dataset.
Human refers to human-rated similarity.}
\label{table:main_results_lexica}
\tabcolsep 3.5pt
\scalebox{0.85}{
\begin{tabular}{l|cccc|c}
\toprule
\textbf{Method} & \textbf{Semantic} & \textbf{Modifier} & \textbf{Image} & \textbf{Pixel} & \textbf{Human}\\
\midrule
ImgCap & 0.19 &  0.00 & 0.65 & 0.89 & 1.65\\
ImgCap (FT) & 0.45 &  0.14 & 0.74 & 0.89 & 3.20\\
CLIP Interrogator & 0.52 & 0.01 & 0.77 & 0.89 & 2.95\\
\attack & \textbf{0.70} & \textbf{0.43} & \textbf{0.80} & \textbf{0.90} & \textbf{4.45}\\
\bottomrule
\end{tabular}
}
\end{table}

\begin{figure}[!t]
\centering
\includegraphics[width=\linewidth]{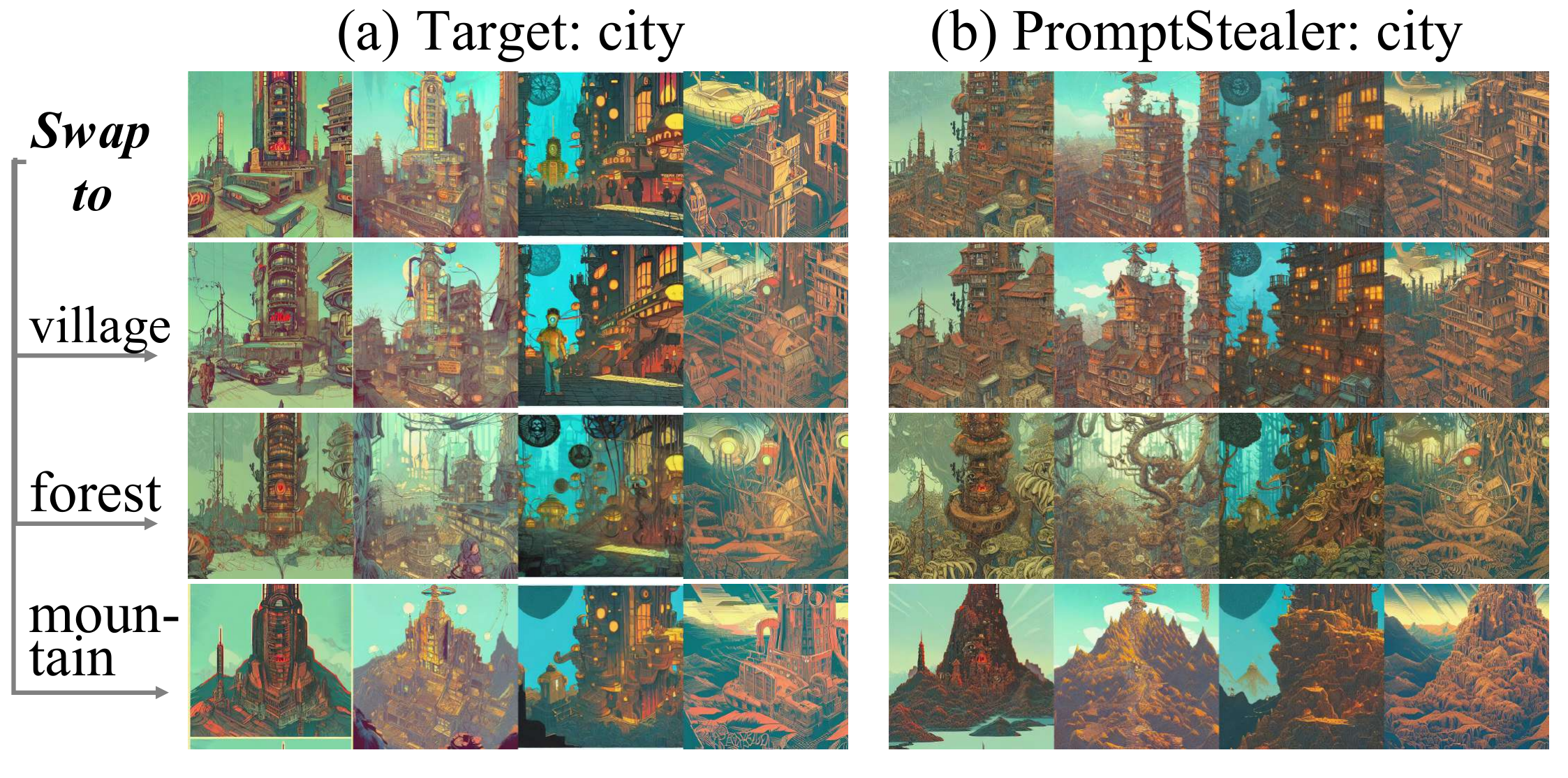}
\caption{Transferability results of the example in \autoref{figure:attack_example_a}.}
\label{figure:transferability_a}
\end{figure}

\begin{figure}[!t]
\centering
\includegraphics[width=\linewidth]{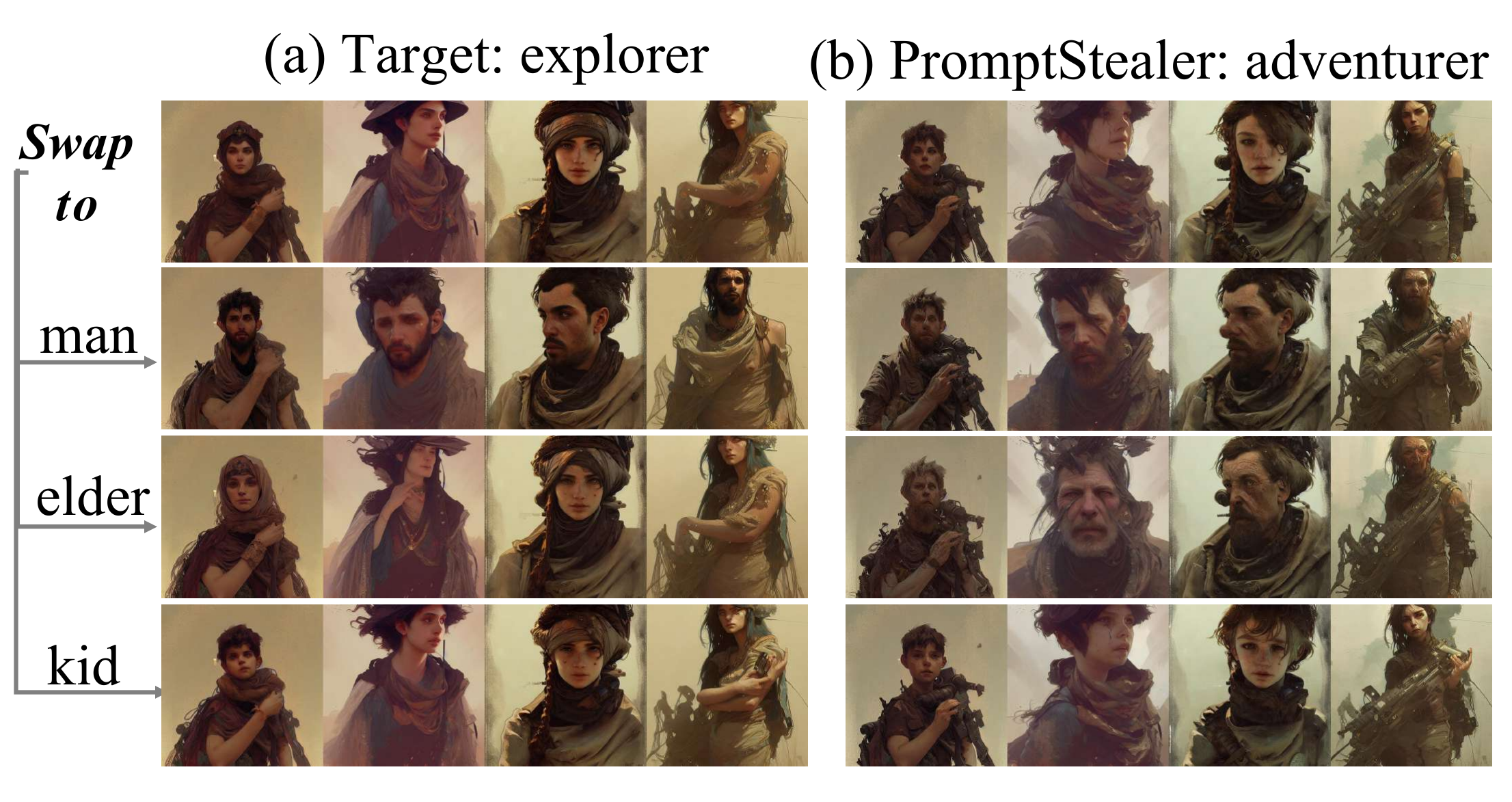}
\caption{Transferability results of the example in \autoref{figure:attack_example_b}.}
\label{figure:transferability_b}
\end{figure}

\subsection{Qualitative Evaluation}

\autoref{table:main_results_lexica} shows the human-rated similarity of \attack and three baselines.
The Cronbach's Alpha among the labelers is 0.90 on average, demonstrating good inter-reliability.
We observe that \attack gets far better human-rated similarity compared with other methods.
For instance, the human-rated similarity for ImgCap, ImgCap (FT), Clip Interrogator, and \attack are 1.65, 3.20, 2.95, and 4.45.

\autoref{figure:attack_exmaples} further shows two attack examples of \attack and the baselines.
Here, for each stolen prompt, we let Stable Diffusion generate four images with different random seeds to eliminate the potential biases.
In both cases, we can see that the stolen images by \attack are more similar to the target images compared to the baselines.
Take \autoref{figure:attack_example_b} as an example, the stolen images by the image captioning method are the least similar to the target image.
Concretely, the first, second, and fourth stolen images by image captioning are in vintage style, while the third image has some modern features with respect to color and clothing fabric.
However, none of the images share a similar style to the target image.
With modifiers introduced, the qualities of the stolen images generated by CLIP Interrogator improved a lot.
All four images show the same vintage yellow tint as the target image.
However, the wrong modifiers steer the stolen images in the wrong direction, thus reducing the efficacy of CLIP Interrogator.
For example, the modifier ``looks like alison brie,'' which is not part of the target prompt, directs the results toward the celebrity Alison Brie (see the first, second, and fourth images).
To compare, we find that \attack is able to recover a large proportion of the modifiers of the target prompt.
The large coverage of modifiers thus promises coherence between the target and stolen images.

As mentioned before, an important usage of the prompt traded through the marketplace is that the user can modify its subject to generate other images with a similar style.
We refer to this usage as transferability. 
Ideally, the stolen prompt should have high transferability as well.
To validate the transferability of the stolen prompts, we continue using two examples in \autoref{figure:attack_exmaples}, and compare the generated images when the subject in the prompt is replaced with new ones.
For example, we replace ``city'' in the subject of the stolen prompt with new ones such as ``village,'' ``forest,'' and ``mountain.''
We also provide the generated images using the target prompts, which are replaced with the same new subjects.
\autoref{figure:transferability_a} and \autoref{figure:transferability_b} showcase the transferability results.
We find that with the stolen prompts, the adversary can generate images that adapt well to other subjects while maintaining similar features to those generated using target prompts, indicating that the stolen prompts by \attack have high transferability.

\subsection{Ablation Study}
\label{section:attack_ablation}

We further evaluate whether \attack is still effective with 1) different subject generators; 2) different modifier detectors; 3) label number; and 4) posterior threshold.

\mypara{Subject Generator}
\autoref{table:ablation_subject_generators} shows the performance of different subject generators.
Regarding the model architectures, we find that BLIP performs slightly better than GIT in the prompt stealing scenario.
Besides, it is clear to observe an increase for both of the two models across all metrics after fine-tuning on \dataset.
For instance, BLIP obtains an increase of 0.90 in human-rated similarity after fine-tuning and exhibits the best performance among all other models.
Therefore, we utilize the fine-tuned BLIP as our subject generator in the following experiments.

\begin{table}[!t]
\centering
\caption{Performances of subject generators.
(FT) denotes that the model is fine-tuned on the subjects and corresponding images of \dataset.}
\label{table:ablation_subject_generators}
\tabcolsep 3.5pt
\scalebox{0.85}{
\begin{tabular}{r|cccc|c}
\toprule
& \textbf{Semantic} & \textbf{Modifier} & \textbf{Image} & \textbf{Pixel} & \textbf{Human}\\
\midrule
GIT & 0.62 & \textbf{0.43} & 0.72 & 0.89 & 3.20\\ 
GIT (FT) & 0.68 & \textbf{0.43} & 0.79 & \textbf{0.90} & 4.25\\ 
BLIP & 0.66 & \textbf{0.43} & 0.79 & \textbf{0.90} & 3.55\\
BLIP (FT) & \textbf{0.70} & \textbf{0.43} & \textbf{0.80} & \textbf{0.90} & \textbf{4.45}\\
\bottomrule
\end{tabular}
}
\end{table}

\mypara{Modifier Detector}
We consider two modifier detector architectures, i.e., ML-Decoder and Query2Label, in our experiments.
As illustrated in \autoref{table:ablation_modifier_detector}, we observe that ML-Decoder achieves better performance than Query2Label, as evidenced by the higher scores across all evaluation metrics.
This can be attributed to the group-decoding scheme of ML-Decoder.
Different from Query2Label which assigns a query per class, ML-Decoder uses a fixed number of queries and interpolates to reach the final number of classes using a group fully-connected block.

\begin{table}[!t]
\centering
\caption{Performances of modifier detectors.}
\label{table:ablation_modifier_detector}
\tabcolsep 3.5pt
\scalebox{0.85}{
\begin{tabular}{r|cccc|c}
\toprule
& \textbf{Semantic} & \textbf{Modifier} & \textbf{Image} & \textbf{Pixel} & \textbf{Human}\\
\midrule
Query2Label & 0.63 & 0.34 & 0.75 & 0.89 & 4.00\\
ML-Decoder & \textbf{0.70} & \textbf{0.43} & \textbf{0.80} & \textbf{0.90} & \textbf{4.45}\\
\bottomrule
\end{tabular}
}
\end{table}

\mypara{Label Number}
In the main experimental setting, we only consider modifiers appearing more than 10 times in \dataset as the labels (7,672) for the multi-label classifier.
We are interested in whether changing the label number affects the attack performance.
To this end, we consider two other variants, i.e., modifiers appear more than 50 times and 100 times which lead to 1,966 labels and 1,109 labels, respectively.
\autoref{figure:label_num} in the Appendix shows the results.
We find that for semantic, modifier, image, and pixel similarities, \attack with 1,109 labels and 1,966 labels have slightly weaker performance than \attack with 7,672 labels.
For instance, the semantic similarities for 1,109, 1,966, and 7,672 labels are 0.67, 0.68, and 0.70, respectively.
This is reasonable as a relatively small modifier set can already cover most modifiers in target prompts, as shown in \autoref{figure:modifier_appearance_times}.

\mypara{Threshold}
\autoref{figure:threshold} in the Appendix shows the results regarding the impact of the posterior threshold for the multi-label classifier.
We find that \attack obtains the highest modifier, image, and pixel similarity when the threshold is set to 0.6.
For semantic similarity, the highest result (0.71) is achieved when the threshold is 0.3, which is very close to the semantic similarity (0.70) when the threshold is 0.6.
In conclusion, 0.6 is a suitable threshold for the multi-label classifier of \attack.

\subsection{Adversary in the Loop}
\label{section:adversary_in_the_loop}

\begin{figure}[!t]
\centering
\includegraphics[width=\linewidth]{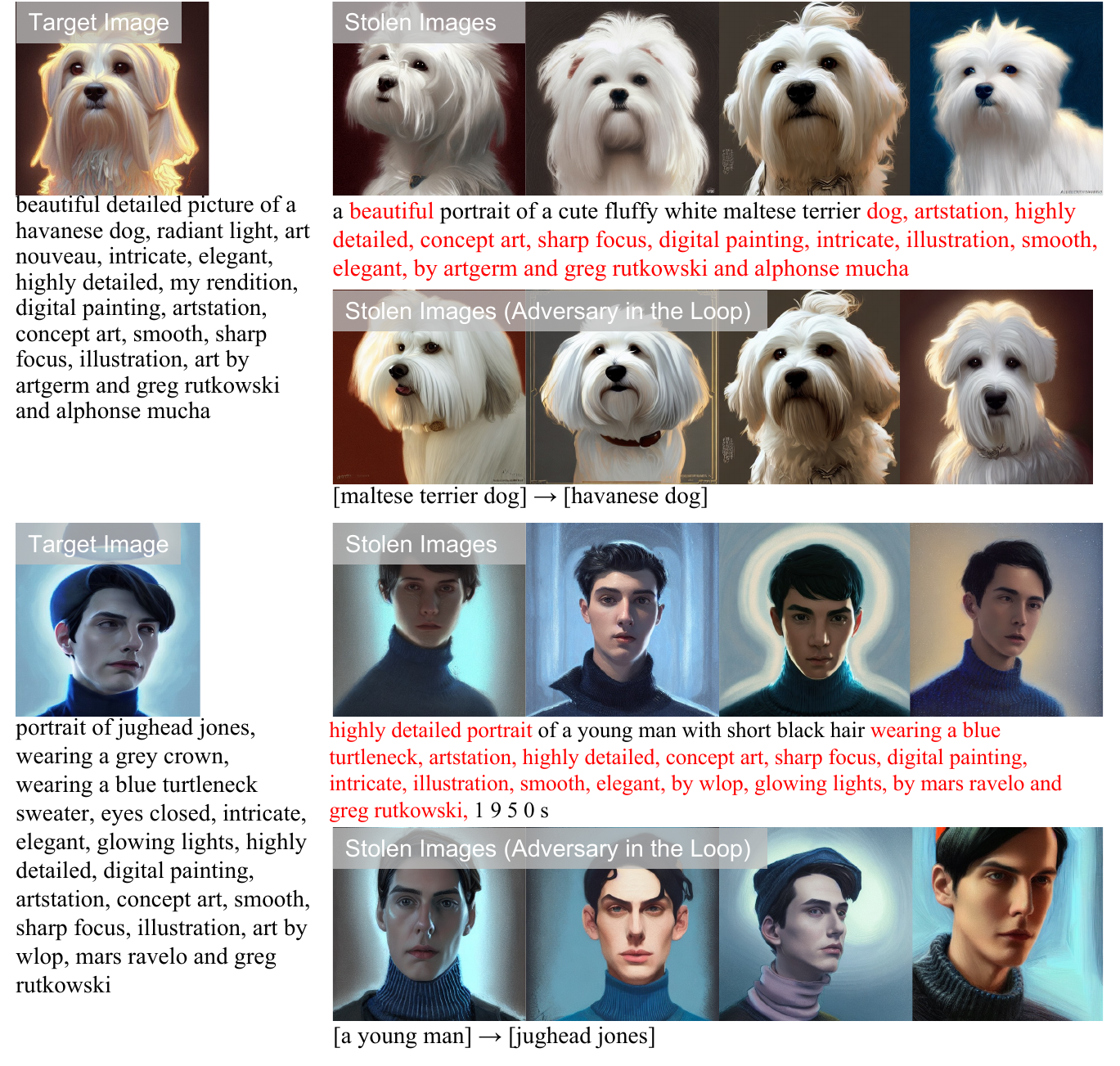}
\caption{Two examples of improving \attack by manually modifying subjects, i.e., adversary in the loop.
The red color marks the correctly predicted modifiers.}
\label{figure:adversary_in_the_loop}
\end{figure}

We acknowledge that \attack is not perfect.
\autoref{figure:adversary_in_the_loop} shows some failed cases of \attack.
We find that the main reason for these failed cases to happen is that \attack cannot accurately capture the key subjects from the target images, such as celebrities and animals.
However, the adversary can easily improve this by manually modifying the stolen prompt with their knowledge, e.g., replacing ``a young man'' with ``jughead jones,'' or modifying the misidentified ``maltese terrier dog'' to ``havanese dog''.
Examples in \autoref{figure:adversary_in_the_loop} show the effects of involving the adversary in the loop.
As we can see, the attack results improve significantly.
Besides involving an adversary in the loop, more advanced machine learning models can be used to solve the issue as well, like adding a subject detector when generating subjects.
We leave this as future work.

\begin{table}[!t]
\centering
\caption{The performance of \attack and baseline methods on DiffusionDB.}
\label{table:main_results_diffusiondb}
\tabcolsep 3.5pt
\scalebox{0.85}{
\begin{tabular}{l|cccc|c}
\toprule
\textbf{Method} & \textbf{Semantic} & \textbf{Modifier} & \textbf{Image} & \textbf{Pixel} & \textbf{Human}\\
\midrule
ImgCap & 0.28 & 0.00 & 0.69 & 0.88 & 1.65\\
ImgCap (FT) & 0.50 & 0.22 & 0.79 & 0.88 & 3.25\\
CLIP Interrogator & 0.51 & 0.01 & 0.79 & 0.88 & 2.15\\
\attack & \textbf{0.64} & \textbf{0.30} & \textbf{0.82} & \textbf{0.89} & \textbf{3.95}\\
\bottomrule
\end{tabular}
}
\end{table}

\begin{table}[!t]
\centering
\caption{The performance of \attack and baseline methods on real-world traded prompts.}
\label{table:main_results_prompt_trading}
\tabcolsep 3.5pt
\scalebox{0.85}{
\begin{tabular}{l|cccc|c}
\toprule
\textbf{Method} & \textbf{Semantic} & \textbf{Modifier} & \textbf{Image} & \textbf{Pixel} & \textbf{Human}\\
\midrule
ImgCap & 0.44 & 0.00 & 0.80 & \textbf{0.89} & 2.85\\
ImgCap (FT) & 0.55 & 0.08 & \textbf{0.83} & 0.88 & 3.35\\
CLIP Interrogator & 0.56 & 0.00 & 0.80 & 0.88 & 3.40\\
\attack & \textbf{0.63} & \textbf{0.22} & \textbf{0.83} & \textbf{0.89} & \textbf{4.05}\\   
\bottomrule
\end{tabular}
}
\end{table}

\begin{figure*}[!t]
\centering
\begin{subfigure}{0.45\linewidth}
\includegraphics[width=\linewidth]{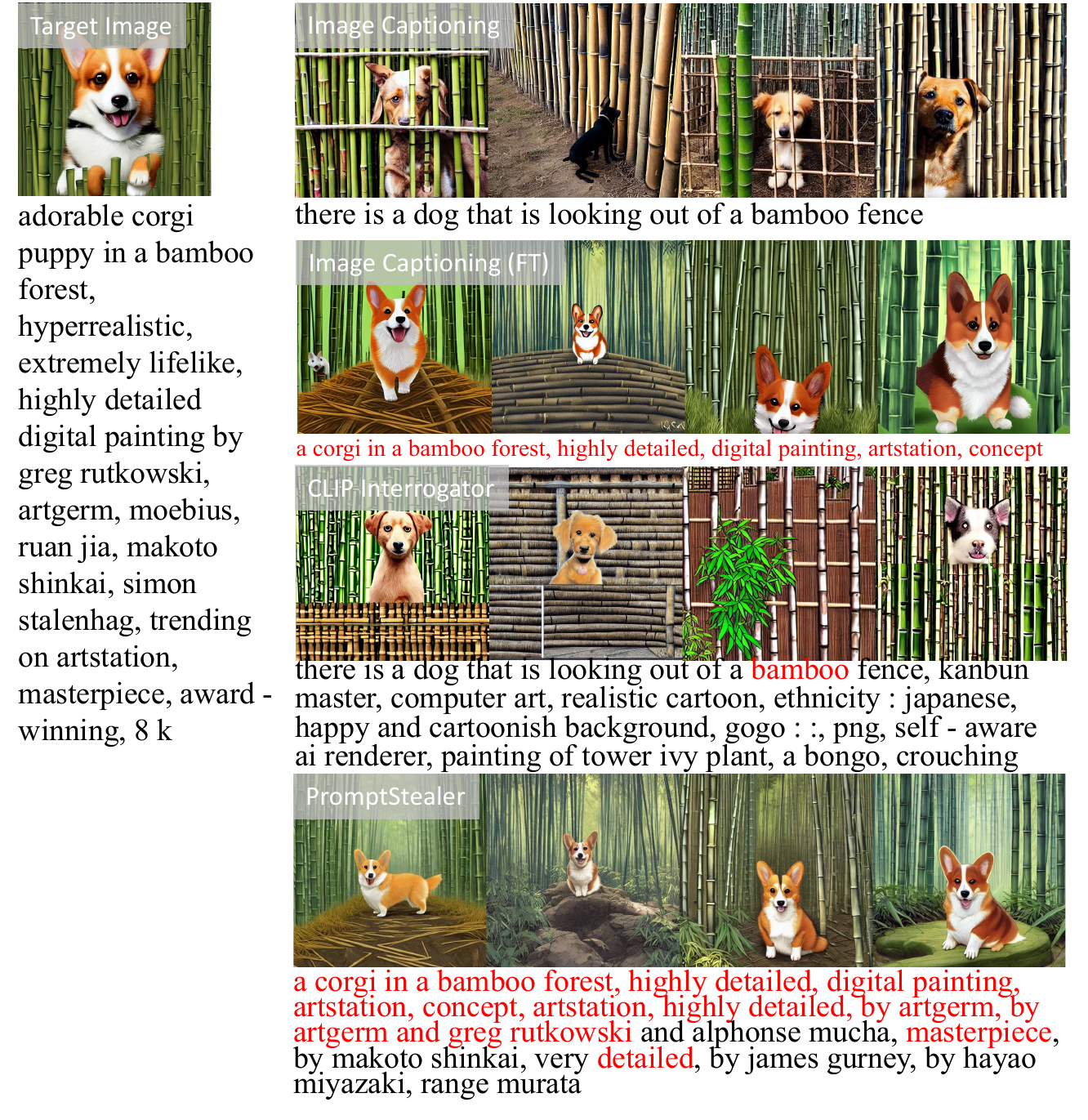}
\caption{}
\end{subfigure}
\begin{subfigure}{0.45\linewidth}
\includegraphics[width=\linewidth]{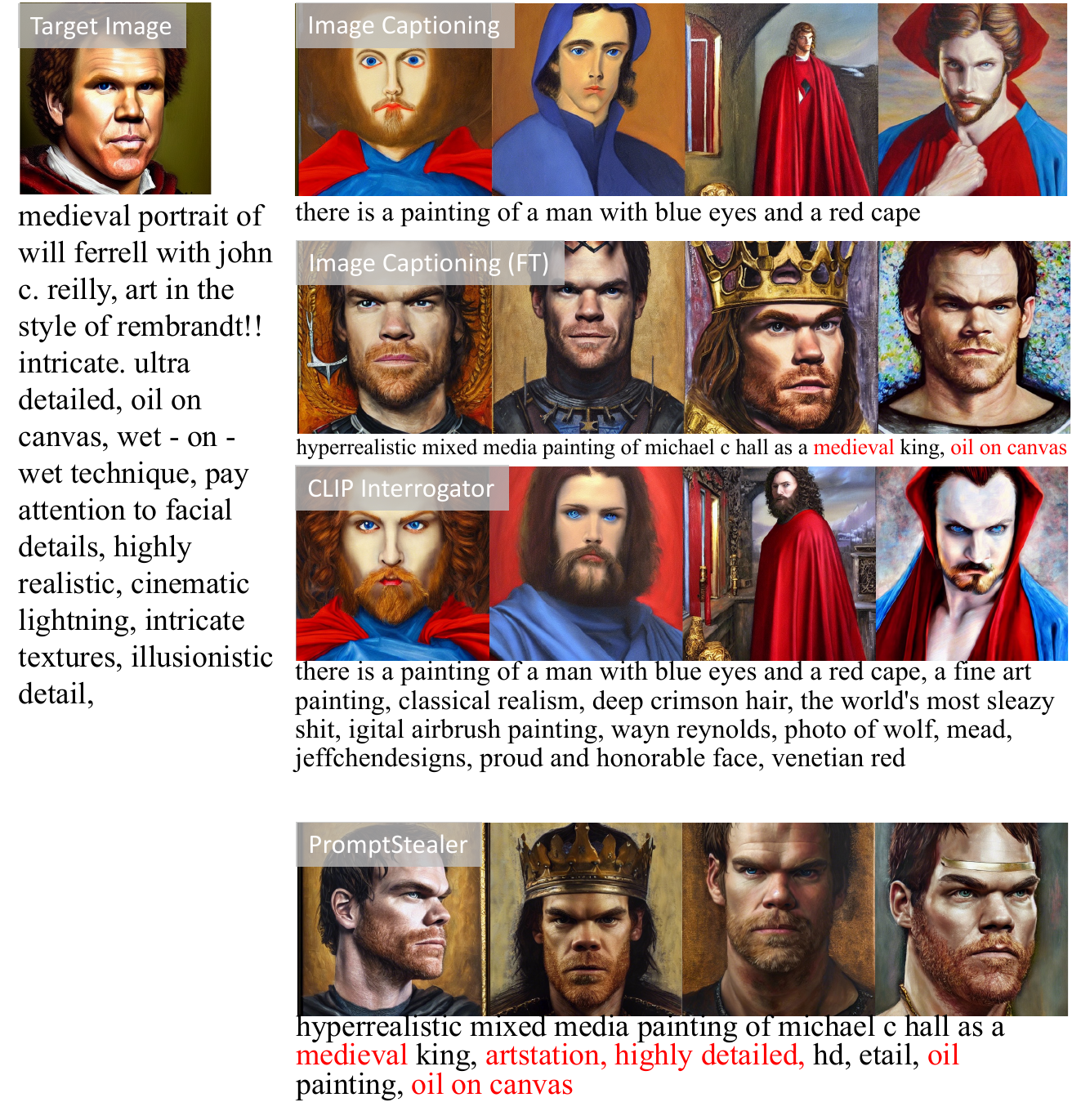}
\caption{}
\end{subfigure}
\caption{Two attack examples of \attack and the three baselines on DiffusionDB.
The text below each image is the target/stolen prompt.}
\label{figure:attack_example_diffusion}
\end{figure*}

\begin{figure*}[!t]
\centering
\begin{subfigure}{0.45\linewidth}
\includegraphics[width=\linewidth]{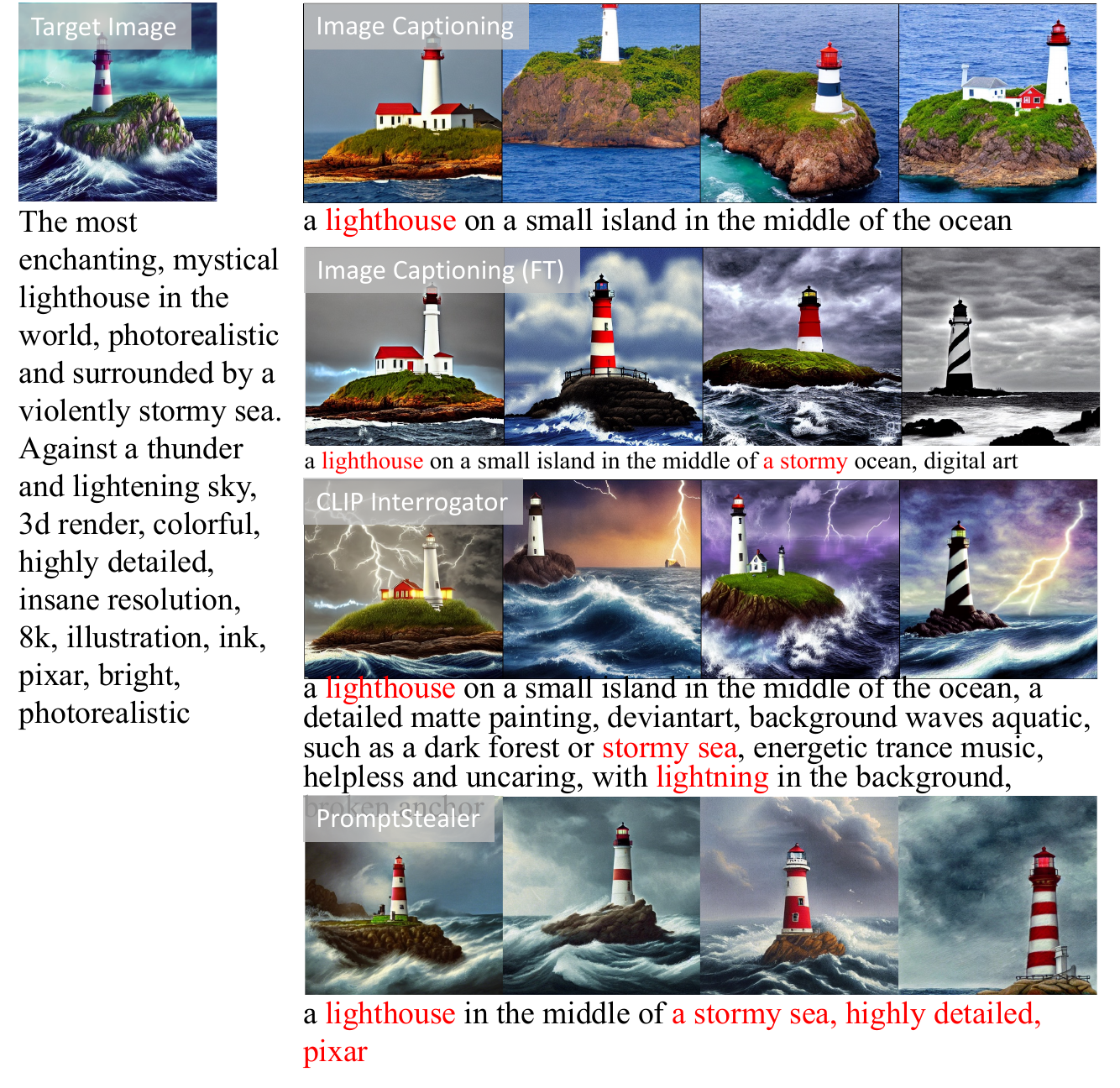}
\caption{}
\end{subfigure}
\begin{subfigure}{0.45\linewidth}
\includegraphics[width=\linewidth]{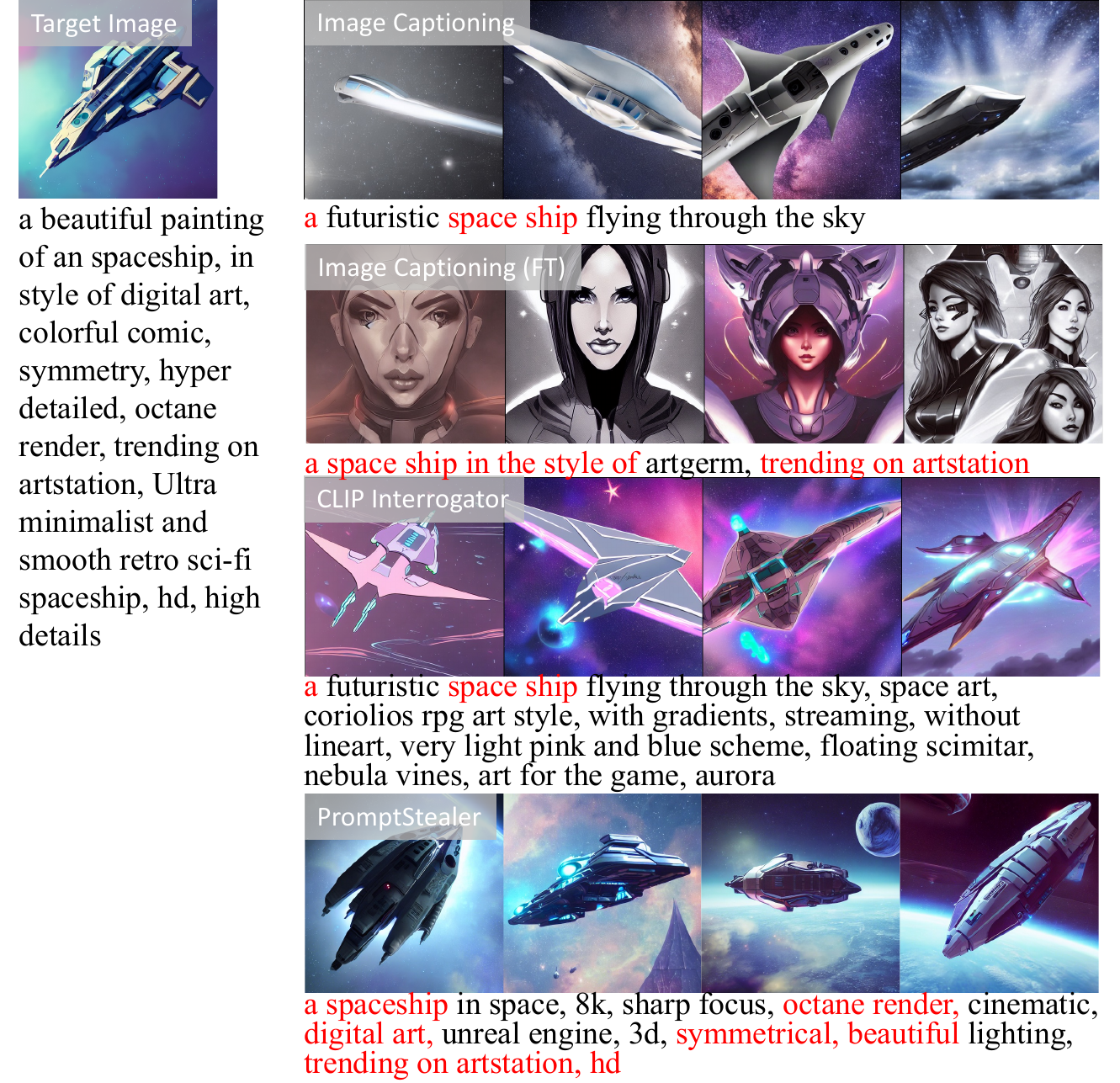}
\caption{}
\end{subfigure}
\caption{Two attack examples of \attack and the three baselines on real-world traded prompts.
The first one is from PromptDB and the latter one is from PromptBase.
The red color marks the correctly predicted modifiers.}
\label{figure:attack_example_marketplace}
\end{figure*}

\subsection{Open-World Evaluation}
\label{section:case_study}

\mypara{Other Datasets}
So far, all the testing samples for \attack are from \dataset.
To simulate more realistic attack scenarios, we evaluate whether \attack can be generalized to target images from out-of-distribution datasets.

\mysubpara{DiffusionDB}
DiffusionDB~\cite{DiffusionDB} is a recent open-source prompt-image dataset and we randomly sample 1k prompts and corresponding images for evaluation.
We find \attack outperforms the three baselines both quantitatively and qualitatively.
As illustrated in \autoref{table:main_results_diffusiondb}, \attack achieves 0.64, 0.30, 0.82, and 0.89 on semantic, modifier, image, and pixel similarity,  respectively.
Regarding qualitative performance, \attack also obtains the highest human-rated similarity with an average score of 3.95.
We display two attack examples in \autoref{figure:attack_example_diffusion}.

\mysubpara{Real-world Traded Prompts}
Additionally, we apply \attack to 10 randomly selected/purchased real-world prompts traded on two prompt marketplaces, PromptBase~\cite{PromptBase} and PromptDB~\cite{PromptDB}.
\autoref{table:main_results_prompt_trading} shows that \attack consistently outperforms baseline methods both quantitatively and qualitatively.
As \autoref{figure:attack_example_marketplace} shows, \attack successfully identifies the main subject and modifiers in the target prompts.
For instance, for the first target image, \attack can successfully infer ``lighthouse'' and ``a stormy sea'' in the subject and the modifiers ``highly detailed'' and ``pixar.''
Moreover, the stolen images exhibit a strong resemblance to the target image, indicating the capability of \attack in open images.

\begin{figure}[!t]
\centering
\includegraphics[width=\linewidth]{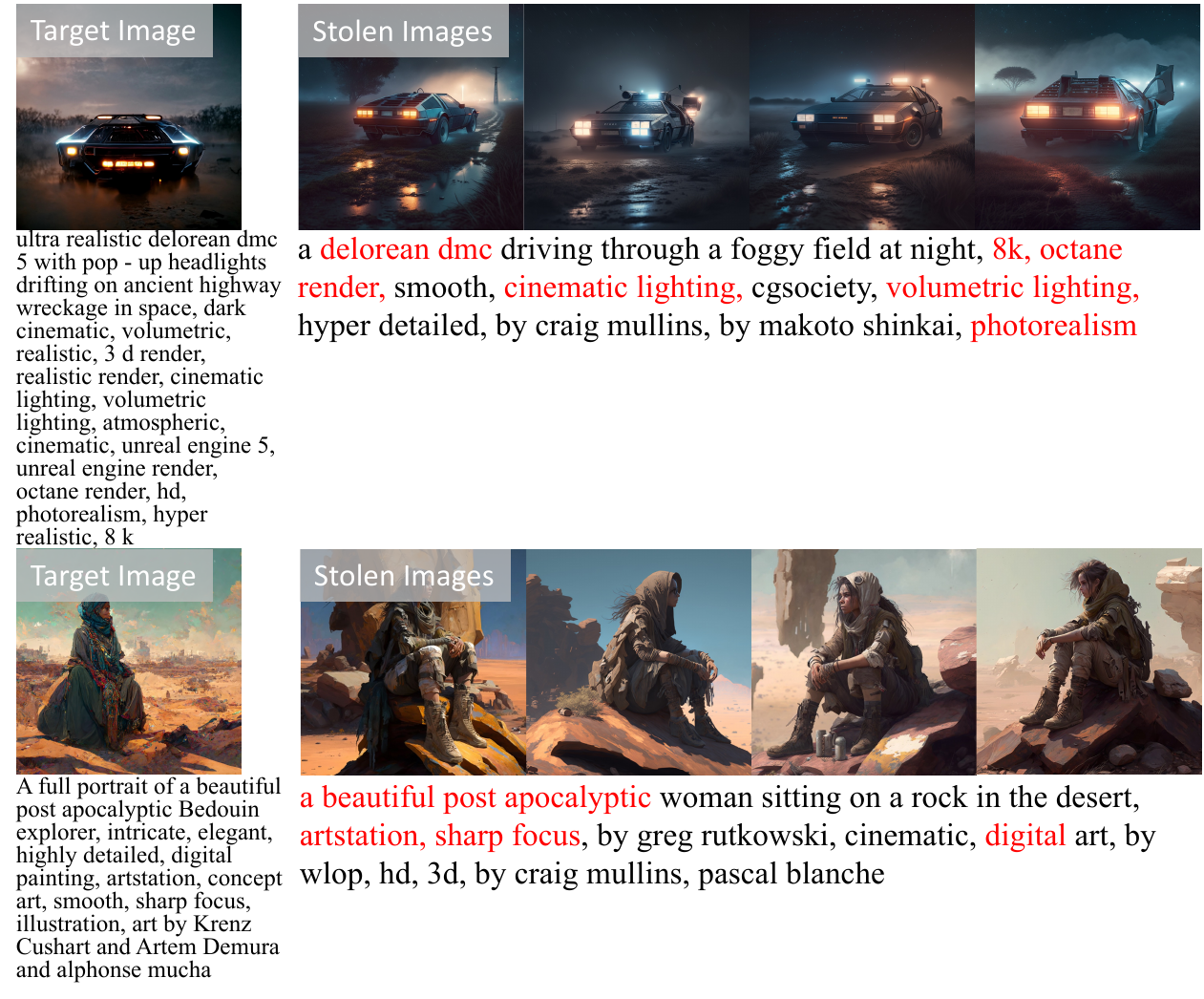}
\caption{Two attack examples of \attack in open-world evaluation (other model Midjourney). 
The target and stolen images are generated by Midjourney.
The red color marks the correctly predicted modifiers.}
\label{figure:midjourney}
\end{figure}

\begin{figure}[!t]
\centering
\includegraphics[width=\linewidth]{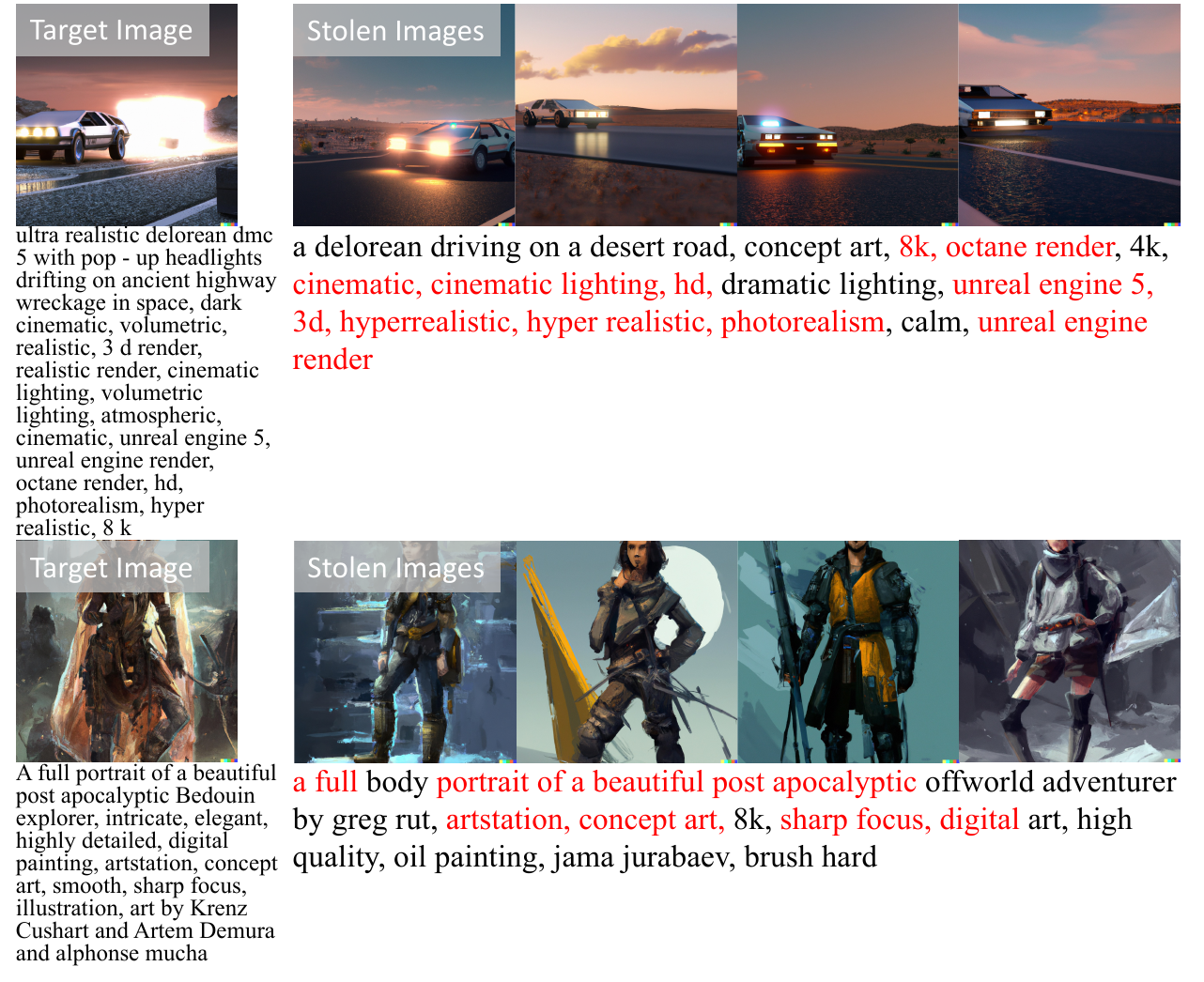}
\caption{Two attack examples of \attack in open-world evaluation (other model DALL$\cdot$E~2). 
The target and stolen images are generated by DALL$\cdot$E~2.
The red color marks the correctly predicted modifiers.}
\label{figure:dalle2}
\end{figure}

\mypara{Other Text-to-Image Generation Models}
Midjourney and DALL$\cdot$E~2 are also mainstream text-to-image generation models.
However, as mentioned before, the two models are still not open-source, so we could not conduct large-scale quantitative experiments on them.
Instead, we perform a case study and hope it could shed light on the generalizability of \attack on these unseen models.
Concretely, we first use two target prompts from \dataset to generate two target images from Midjourney and DALL$\cdot$E~2.
We then directly apply the trained \attack to steal the prompts from the two target images and use the corresponding text-to-image generation models again to get the stolen images.
The results are displayed in \autoref{figure:midjourney} and \autoref{figure:dalle2}.

We find that even though \attack has never seen images generated from Midjourney and DALL$\cdot$E~2 before, it still manages to catch certain key prompt modifiers, such as ``octane render,'' ``artstation,'' ``volumetric lighting.''
This indicates that \attack still works on these unseen text-to-image generation models to a certain extent.
We also notice that \attack does not perform as well as it does on Stable Diffusion.
For example, it cannot accurately deduce the artist modifiers.
This could be caused by the different architectures and weights of these text-to-image generation models. 
One possible solution is to train \attack on multiple datasets related to different models like Midjourney and DALL$\cdot$E~2.
However, as no dataset is available by the time we perform this study, we leave this as future work.

\mypara{Newer Versions of Stable Diffusion}
With the advancement of Stable Diffusion, it is also interesting to observe whether \attack works on different versions of Stable Diffusion.
Note that the data we collected contains no model version information.
Therefore, given a target prompt, we utilize a certain version of Stable Diffusion to generate the target image and then apply \attack to steal its prompts.
We test three versions of Stable Diffusion: v1.4, v1.5, and v2.0.
\autoref{table:different_sd_versions} shows the results.
Overall, \attack performs similarly on v1.4 and v1.5, but slightly worse on v2.0.
This may be because v2.0 came out after our data collection, so its samples are not included in the training set.
We also display one attack example in \autoref{figure:sd_version} in the Appendix.

\begin{table}[!t]
\centering
\caption{Performances across Stable Diffusion versions.}
\label{table:different_sd_versions}
\tabcolsep 3.5pt
\scalebox{0.85}{
\begin{tabular}{c|cccc|c}
\toprule
\textbf{Version} & \textbf{Semantic} & \textbf{Modifier} & \textbf{Image} & \textbf{Pixel} & \textbf{Human} \\
\midrule
v1.4 & \textbf{0.63} & 0.43 & 0.79 & 0.90 & \textbf{4.5}\\
v1.5 & \textbf{0.63} & \textbf{0.45} & 0.79 & \textbf{0.91} & 4.3\\
v2.0 & 0.53 & 0.32 & \textbf{0.82} & 0.88 & 4.25\\
\bottomrule
\end{tabular}
}
\end{table}

\subsection{Performance on Multiple Target Images}

In some cases, the adversary is able to obtain multiple target images generated by one target prompt.
Therefore, we also explore the potential of \attack when multiple target images are provided.
In this scenario, the adversary first gets a set of the stolen prompts (including the subject and modifiers) via \attack and then applies different strategies to get the best prompts. 
We examine three strategies: 
1) random: randomly choose one from all stolen prompts; 2) best: choose the stolen prompt achieving the best performance; 3) union: greedy-search the best caption and union all modifiers predicted from these target images.
The quantitative and qualitative results can be found in \autoref{table:multiple_image_quantitative} and \autoref{figure:multi_target} in the Appendix.
We find that providing multiple target images indeed increases the attack performance, and the union strategy performs the best. 

\begin{table}[!t]
\centering
\caption{Performances of different strategies when facing multiple target images. ``Single'' is the performance of a single target image.}
\label{table:multiple_image_quantitative}
\tabcolsep 3.5pt
\scalebox{0.85}{
\begin{tabular}{r|cccc|c}
\toprule
& \textbf{Semantic} & \textbf{Modifier} & \textbf{Image} & \textbf{Pixel} & \textbf{Human}\\
\midrule
Single & 0.70 & 0.43 & 0.80 & 0.90 & 4.45\\
\midrule
Random & 0.70 & 0.43 & 0.80 & 0.90 & 4.25\\
Best & 0.72 &  0.42 & \textbf{0.83} & 0.90 & 4.50\\
Union & \textbf{0.77} & \textbf{0.44} & \textbf{0.83} & \textbf{0.91} & \textbf{4.65}\\
\bottomrule
\end{tabular}
}
\end{table}

\subsection{ChatGPT in Prompt Stealing Attacks}

\begin{figure}[!t]
\centering
\includegraphics[width=\linewidth]{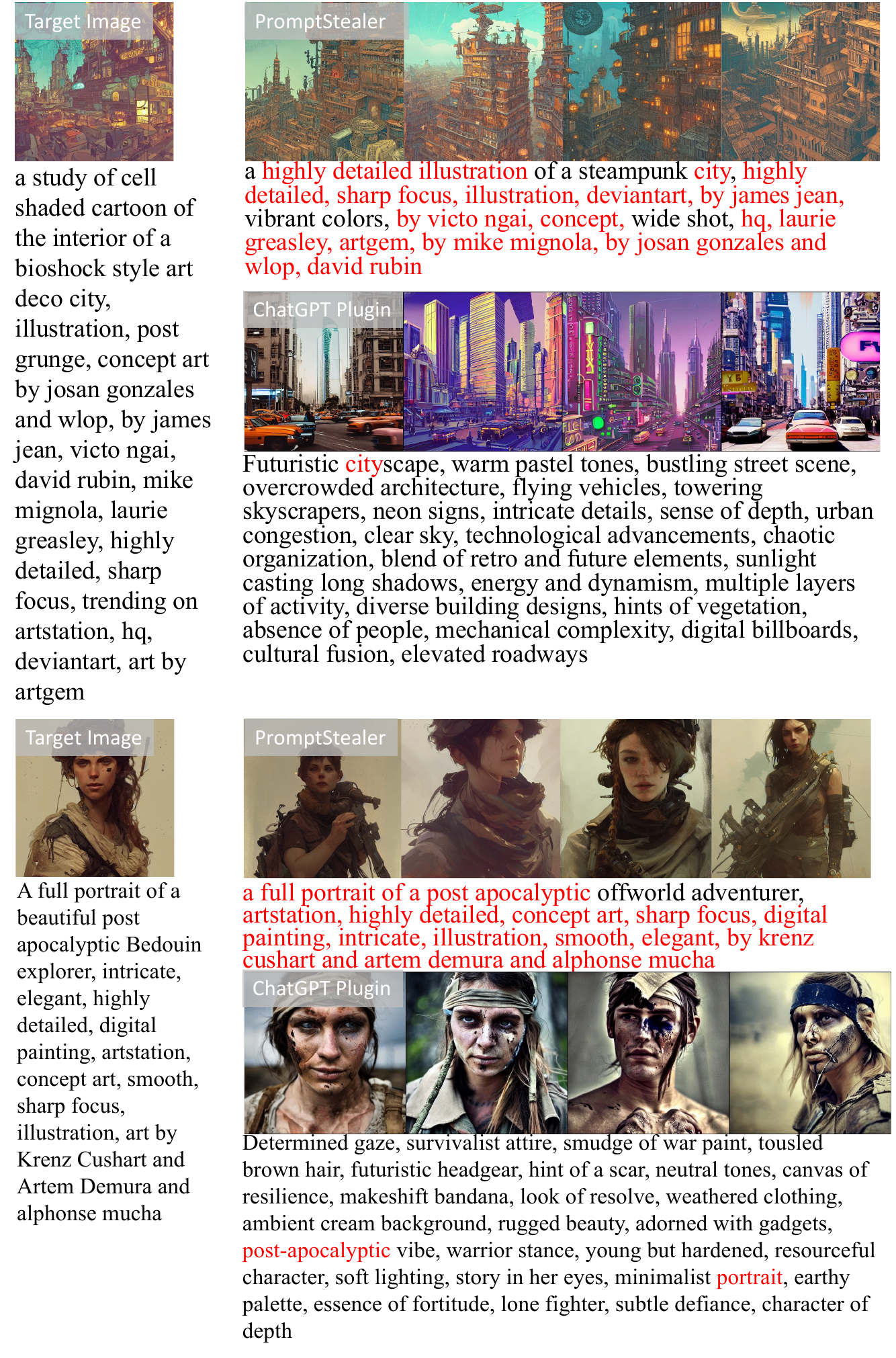}
\caption{Two attack examples of \attack and ChatGPT Plugins.
The red color marks the correctly predicted modifiers.}
\label{figure:chatgpt}
\end{figure}

ChatGPT (GPT-4), as an advanced vision-language model, can also be prompted to infer the target prompt given a target image.  
Therefore, we evaluate its performance by leveraging a prevalent prompt generator in the GPT store~\cite{PromptGenerator}.
Considering the required manual efforts, we randomly pick ten test samples in \dataset.
Its semantic, modifier, image, pixel, and human-rated similarities are 0.35, 0.00, 0.72, 0.91, and 3.2, suggesting \attack still outperforms ChatGPT.
In \autoref{figure:chatgpt}, we present two attack examples between \attack and ChatGPT.
A key observation is that \attack outperforms ChatGPT in recognizing the art style of the target image.
For instance, in the first example, while ChatGPT accurately recognizes the subject as a city, it incorrectly identifies the target art style - from ``cartoon, Josan Gonzales, Wlop'' to ``Futuristic, warm pastel tones.''
This misinterpretation significantly hampers the quality of the output.

\section{Defense}
\label{section:defense}

After demonstrating the efficacy of \attack, we further propose a defense method, namely \emph{\defense}.

\subsection{\defense}

The adversary relies on machine learning models to realize \attack.
To mitigate the attack, one natural solution is to use adversarial examples to reduce the performance of \attack's machine learning models.
Specifically, we aim to add an optimized perturbation to a target image to obtain a \emph{shielded image} such that \attack cannot infer the stolen prompt from the shielded image effectively.
\defense has two goals, i.e., effectiveness and utility.
High effectiveness indicates the performance of \attack drops significantly on shielded images; high utility implies perturbations on the shielded images are imperceptible to humans.

\attack consists of two machine learning models: a subject generator and a modifier detector.
Technically, \defense can generate shielded images against either of the models or both.
Here, we focus on the modifier detector instead of the subject generator.
The reason is that if \defense creates a shielded image to mislead the subject generator, then \attack will obtain a subject misaligned with the shielded image.
By manually checking, the adversary can easily fix the error in the subject, as we show in \autoref{section:adversary_in_the_loop}, which reduces the effectiveness of \defense. 
Meanwhile, if the shielded image is against the modifier detector, then \attack will infer a stolen prompt with wrong or missing modifiers.
As each prompt has multiple modifiers and the total number of modifiers is large, it is not easy for the adversary to spot which modifiers are missing or wrong.

To generate a shielded image against the modifier detector, \defense can choose any or all of the 7,672 labels/modifiers. 
Here, we choose the modifiers in the artist category.
The reasons are twofold.
First, as shown in \autoref{section:data_analysis}, artist modifiers play important roles in driving the generated images to specific styles. 
Second, focusing on artist modifiers also directly protects artists' intellectual properties.

The concrete process of \defense is as follows.
For the modifier set of a target prompt, we first remove all its artist modifiers to obtain a shielded modifier set.
Then, we optimize a perturbation for the target image such that the final shielded image is classified towards the shielded modifier set.
In other words, we do not mislead the modifier detector to classify the shielded image to a different set of artist modifiers.
Instead, we generate the perturbation that can ideally remove the artist-related information from the target image.
For optimization, we adopt I-FGSM~\cite{KGB16}.
We set the iterative step to 100 and $\epsilon$ to 0.2.
We also evaluate the performance of another representative method, namely C\&W~\cite{CW17}.
The performance is reported in \autoref{section:defense_cw}.
As the design principle of \defense is general, we further apply it to the optimization-based baseline attack CLIP Interrogator.
The results are listed in \autoref{section:defense_clip_interrogator}.
Note that we assume the defender has white-box access to the modifier detector of \attack.
We acknowledge that this is a strong assumption and the main weakness of our defense.

\subsection{Experimental Settings}
\label{section:defense_exp_settings}

We follow the same experimental settings in \autoref{section:exp_settings} for the text-to-image generation model and the attack model.

\mypara{Evaluation Metric}
We design our evaluation metrics based on the two goals of the defender: effectiveness and utility.
\begin{itemize}
\item \mypara{Effectiveness} 
We adopt the same quantitative metrics, i.e., semantic, modifier, image, and pixel similarity, from \autoref{section:exp_settings} as the effectiveness metrics. 
\item \mypara{Utility} 
For utility, we measure the mean squared error (MSE) between the target image and the shielded image. 
Lower MSE implies higher utility.
\end{itemize}

Besides, we also provide human-rated similarity as qualitative evaluation.

\begin{table}[!t]
\centering
\caption{The performance of \defense against \attack.
The second row refers to modifier categories.}
\label{table:defense_for_promptstealer}
\tabcolsep 3.5pt
\scalebox{0.85}{
\begin{tabular}{l|ccccc}
\toprule
& \textbf{Semantic} & \textbf{Modifier} & \textbf{Image} & \textbf{Pixel} & \multicolumn{1}{|c}{\textbf{Human}}\\
\midrule
Unshielded & 0.70 & 0.43 & 0.80 & 0.90 & \multicolumn{1}{|c}{4.45}\\
Shielded & 0.62 & 0.74 & 0.71 & 0.88 & \multicolumn{1}{|c}{1.85}\\
\midrule
& \textbf{Artist} & \textbf{Medium} & \textbf{Flavor} & \textbf{Movement} & \textbf{Trending}\\
\midrule
Unshielded & 0.49 & 0.44 & 0.43 & 0.03 & 0.42\\
Shielded & 0.06 & 0.53 & 0.79 & 0.06 & 0.48\\
\bottomrule
\end{tabular}
}
\end{table}

\subsection{Quantitative Evaluation}
\label{section:defense_promptstealer_quantitative}

\autoref{table:defense_for_promptstealer} shows the effectiveness of \defense.
We observe that the artist modifiers exhibit the greatest reduction in similarity (from 0.49 to 0.06) while the overall modifier similarity increases.
This is expected, as the shielded image contains less information related to artist modifiers, the attack model has a higher capacity to predict other modifiers more accurately.
On the other hand, we find that both semantic and image similarities decrease.
Concretely, the semantic similarity decreases from 0.70 to 0.62, and the image similarity decreases from 0.80 to 0.71.
Considering the previous worse baseline (the image captioning method) only gets a 0.65 image similarity, we conclude that \defense achieves strong performance in defending against \attack.

Regarding utility, we find \defense performs well in producing concealed perturbation.
The average MSE between the target and shielded images is 0.0007 which implies that the added perturbation is imperceptible to humans.

We observe comparable results when C\&W is adopted as the optimization method for \defense (see \autoref{section:defense_cw}).
In addition, \defense also achieves promising results on CLIP Interrogator (see \autoref{section:defense_clip_interrogator}).

\begin{figure}[!t]
\centering
\includegraphics[width=\linewidth]{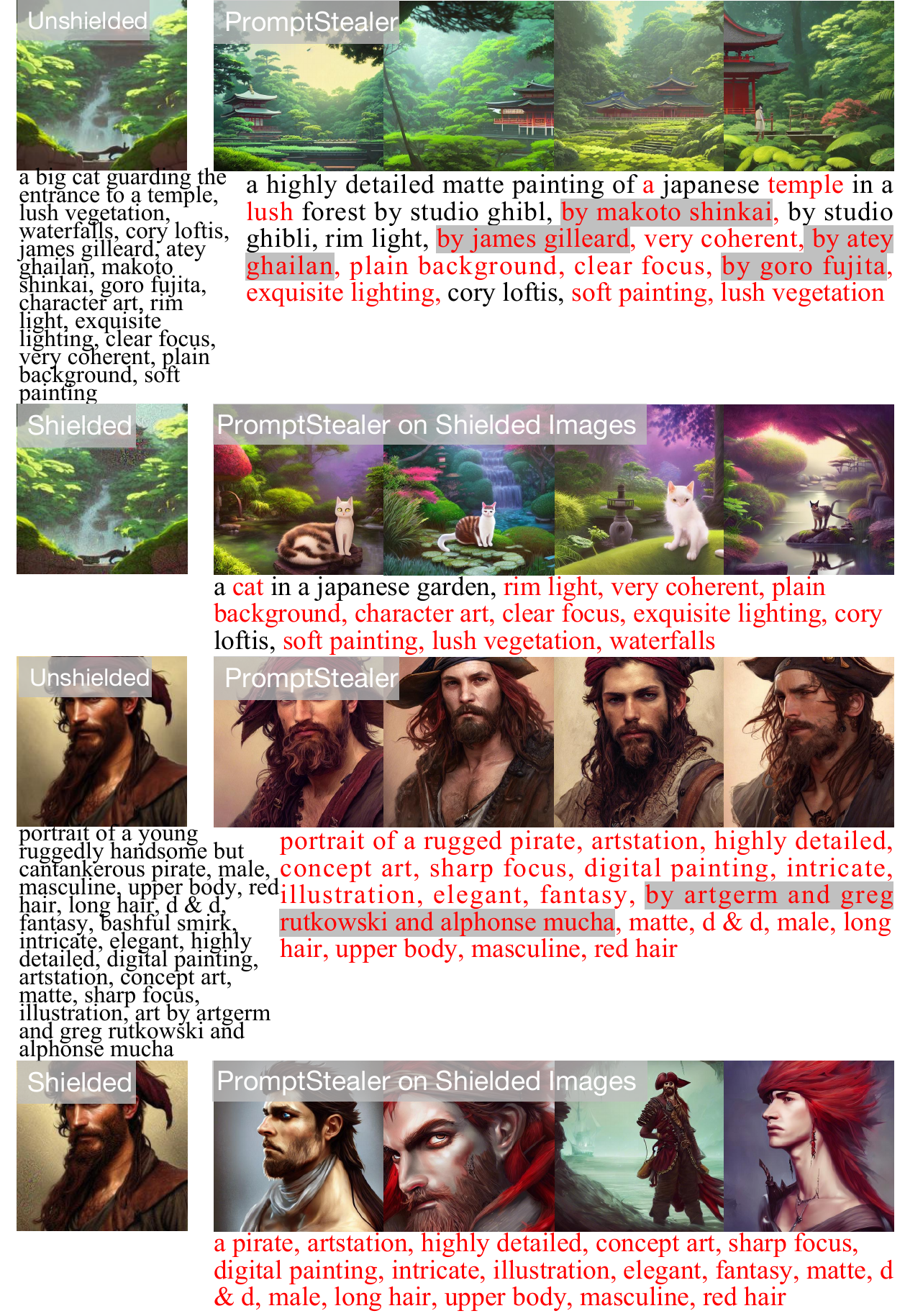}
\caption{Two defense examples of \defense against \attack on the artist modifiers.
The red color marks the correctly predicted modifiers. 
We highlight in grey the target artist modifiers stolen by \attack.}
\label{figure:defense_examples_promptstealer}
\end{figure}

\subsection{Qualitative Evaluation}

\autoref{figure:defense_examples_promptstealer} further shows two defense examples of \defense against \attack.
In both cases, we can see that the stolen images originating from the shielded images are less similar to the target images.
Take the first target image in \autoref{figure:defense_examples_promptstealer} as an example.
The artist modifiers in the target prompt is ``james gilleard,'' ``atey ghailan,'' ``makoto shinkai,'' and ``goro fujita.''
We find that for the unshielded target image, the artist modifier can be successfully stolen by \attack.
However, after applying \defense, the attack model cannot predict the artist modifier from the shielded image.
Moreover, even though the stolen prompt based on the shielded image contains some correct modifiers such as ``lush vegetation,'' the stolen images are quite different from the target image without the artist modifier.
This has been further confirmed in our human study.
As shown in \autoref{table:defense_for_promptstealer}, the human-rated similarity decreases from 4.45 to 1.85.
Besides, by comparing the target and shielded images, we find they are quite similar, demonstrating the utility of \defense.

\subsection{Limitations}

As mentioned before, \defense is effective but it requires a strong assumption for the defender, i.e., white-box access to the attack model.
We have also experimented with the transfer defense.
Concretely, we apply \defense on CLIP Interrogator and use the generated perturbation to defend \attack; however, the experimental results are not promising.
In addition, our evaluation shows that the defense performance can be reduced by the adaptive attack, which trains the modifier detector with shielded images and ground-truth modifiers (see \autoref{table:defense_adaptive} in Appendix).
We emphasize that our goal is to assess whether the defense under a strong assumption is effective, and we hope our results can provide guidance for developing more advanced defenses in the future.

\section{Related Work}

\mypara{Prompt Engineering}
Prompt engineering in text-to-image generation models aims to improve the quality of generated images by developing prompt design guidelines~\cite{PU22,LC22,O22}.
Liu and Chilton~\cite{LC22} qualitatively investigate what prompt components and model parameters can produce high-quality images and provide seven suggestions for prompt design.
They emphasize that using proper modifiers, rather than rephrasing the prompt with the same modifier set, is a key factor in image generation quality.
Oppenlaender~\cite{O22} presents a taxonomy of prompt modifiers based on an ethnographic study and highlights the significance of correctly using prompt modifiers.
Pavlichenko and Ustalov~\cite{PU22} study a human-in-the-loop approach to find the most useful combination of prompt modifiers with a genetic algorithm.
Overall, these works suggest that modifiers, especially proper modifiers, are crucial in generating high-quality images.
In addition to our data analysis in \autoref{section:data_analysis}, these previous works also inspire the design of \attack.

\mypara{Image-to-Text Synthesis}
There are also studies focusing on image-to-text synthesis, such as image captioning tasks~\cite{LLXH22, ClipInterrogator, MHB21}.
Li et al.\ adopt the multimodal mixture of encoder-decoder architecture and an additional filter to remove the noisy captions during the caption generation process~\cite{LLXH22}.
Wang et al.\ design GIT, which simplifies the image captioning architecture to an image encoder and a text decoder.
Besides, CLIP Interrogator, a tool in the open-source community also put efforts into searching prompts to match a given image~\cite{ClipInterrogator}.
CLIP Interrogator relies on CLIP to iteratively calculate the similarity between modifiers and the target image.
It is inefficient due to its iterative design and many manually defined hyperparameters.
Intuitively, the above-mentioned works can all be tailored to steal prompts.
However, our experimental results reveal these methods fail to reverse prompts satisfactorily. 
Instead, we propose a simple yet effective prompt stealing attack.

\mypara{Security Issues of Machine Learning Models}
The security issues of machine learning models have been increasingly discussed for years.
Researchers have identified various types of attacks such as adversarial examples~\cite{CW17, LV15, PMGJCS17, PMJFCS16, CJW20}, membership inference~\cite{SSSS17, SZHBFB19, NSH19, SSM19, LZ21, CTCP21, CCNSTT22}, backdoor~\cite{GDG17, LMALZWZ18, YLZZ19, SSP20, LMBL20, SHLSBZ22, JLG22}, etc.
Most of these works focus on classification models.
The security vulnerabilities of text-to-image generation models have received little attention.
To the best of our knowledge, only limited works discuss the possible attack vector or misuse of the text-to-image generation models~\cite{SCWZHZ23, WYLBZ22, LPNDTT23, CHNJSTBIW23, QSHBZZ23}.
However, none of them investigates the emerging threat of stealing prompts from the images.
We hope our study can provide insights into this novel attack and inspire more research in this field.

\section{Discussion}

\mypara{Legal Debates / Copyright Laws Concerning Prompts}
The legal environment for AI-generated content (AIGC) is quickly changing, with various legal authorities worldwide working to adapt copyright laws for the digital era.
In 2023, the US Copyright Office opened a public comment period from August to October to address complex issues related to AI and copyright~\cite{us_aigc_copyright}.
Currently, copyright ownership in AIGC remains ambiguous due to unclear data collection regulation, the need for equitable benefit distribution, inconsistent global legal views on AI copyright, and challenges in tracing all original works used in AI training~\cite{L23}.
Compared to the complex copyright landscape of generated images, user-curated prompts involve significant human efforts and thereby align more closely with traditional copyright concepts that copyright laws protect works created and fixed by humans~\cite{us_copyright, prompt_copyright}.
Prompt marketplaces like PromptBase and PromptSea also explicitly state that the prompts on their platforms are owned by the individuals who created them~\cite{PromptBase_terms_of_service, PromptSea_whitepaper}.
Therefore, prompt stealing attacks infringe on intellectual property rights.

\mypara{Societal Impacts}
This work has three important implications for the AIGC participants, prompt engineers, and marketplace owners.
First, our work demonstrates the feasibility of prompt stealing attacks.
Given the image showcased in the prompt marketplace, an adversary is capable of inferring the selling prompt without payment quickly.
The extremely short attack time makes the threat severity of prompt stealing attacks more than just stealing a certain prompt, but represents a new data breach channel to prompt marketplaces.
For example, an adversary can perform prompt stealing attacks to quickly steal thousands of prompts on marketplaces and sell them on underground forums or a competitive marketplace.
This data breach incident causes substantial financial losses to the victim marketplaces and jeopardizes their business models, which has also been seriously discussed in PromptSea’s white paper~\cite{PromptSea_whitepaper}.
Second, we bring insights on how to mitigate prompt stealing attacks, i.e., by introducing unperceived perturbations on the showcased images.
We argue that there is an urgency to propose effective and flexible defenses against prompt stealing attacks.
The proposed defense method \defense can serve as a foundational baseline for this direction.
Third, we contribute by collecting \dataset, a dataset with 61,467 prompt-image pairs and categorized modifiers.
This dataset can be used not only in training the attack model but also to provide insights to prompt engineers such as helping them to understand the impact of a certain modifier or the joint effects among modifier combinations.
As the first systematic study of prompt stealing attacks, we believe that our findings can serve as a valuable resource for the research community and stakeholders to navigate and mitigate this emerging threat.

\mypara{Limitations \& Future Work}
Our work has limitations.
First, we mainly evaluate the attack on Stable Diffusion and perform case studies on other text-to-image models such as Midjourney and DALL$\cdot$E~2.
Though \attack infers certain keywords under this open-world setting, it is essential to quantitatively evaluate whether other text-to-image models are under the threat of prompt stealing attacks.
Considering Midjourney and DALL$\cdot$E~2 are close-sourced, we leave this as future work. 
Second, our research focuses on prompts for text-to-image generation models.
But as prompts for large language models are appearing on marketplaces, it would be interesting to investigate if prompt stealing attacks also apply to them.

\section{Conclusion}

In this paper, we conduct the first investigation on prompt stealing attacks against text-to-image generation models.
A successful prompt stealing attack directly violates the intellectual property of prompt engineers and threatens the business model of prompt marketplaces.
In detail, we first collect a dataset \dataset and perform a large-scale measurement on it to show that a successful prompt stealing attack should consider a target prompt's subject as well as its modifiers.
We then propose a simple yet effective prompt stealing attack, \attack.
Experimental results show that \attack outperforms the three baseline methods both quantitatively and qualitatively.
We also make attempts to defend against \attack.
In general, our study shed light on the threats existing in the ecosystem created by the popular text-to-image generation models.

\medskip
\mypara{Acknowledgments}
We thank all anonymous reviewers for their constructive comments. This work is partially funded by the European Health and Digital Executive Agency (HADEA) within the project “Understanding the individual host response against Hepatitis D Virus to develop a personalized approach for the management of hepatitis D” (DSolve) (grant agreement number 101057917).

\begin{small}
\bibliographystyle{plain}
\bibliography{normal_generated_py3}    
\end{small}

\appendix
\section*{Appendix}

\section{The Impacts of Modifier Counts on CLIP Interrogator}
\label{section:clip_inter_modifiers}

As illustrated in \autoref{table:label_impact_clip_interrogator}, changing modifier counts from 77,616 to 7,637 resulted in minimal variations for CLIP Interrogator in semantic, modifier, image, and pixel similarity, with changes of -0.007, +0.007, -0.014, and +0.000, respectively.
Therefore, we utilize CLIP Interrogator with 77,616 modifiers in comparison, since it achieves slightly better performance.

\begin{table}[ht]
\centering
\caption{The impacts of modifier counts on CLIP Interrogator.}
\label{table:label_impact_clip_interrogator}
\tabcolsep 3.5pt
\scalebox{0.85}{
\begin{tabular}{r|cccc}
\toprule
\textbf{\# Modfiers} & \textbf{Semantic} & \textbf{Modifier} & \textbf{Image} & \textbf{Pixel}\\
\midrule
77,616 & \textbf{0.52} & 0.01 & \textbf{0.77} & \textbf{0.89}\\
7,637 & 0.51 & \textbf{0.02} & 0.75 & \textbf{0.89}\\ 
\bottomrule
\end{tabular}
}
\end{table}

\section{\defense With C\&W}
\label{section:defense_cw}

\begin{figure}[!t]
\centering
\includegraphics[width=\linewidth]{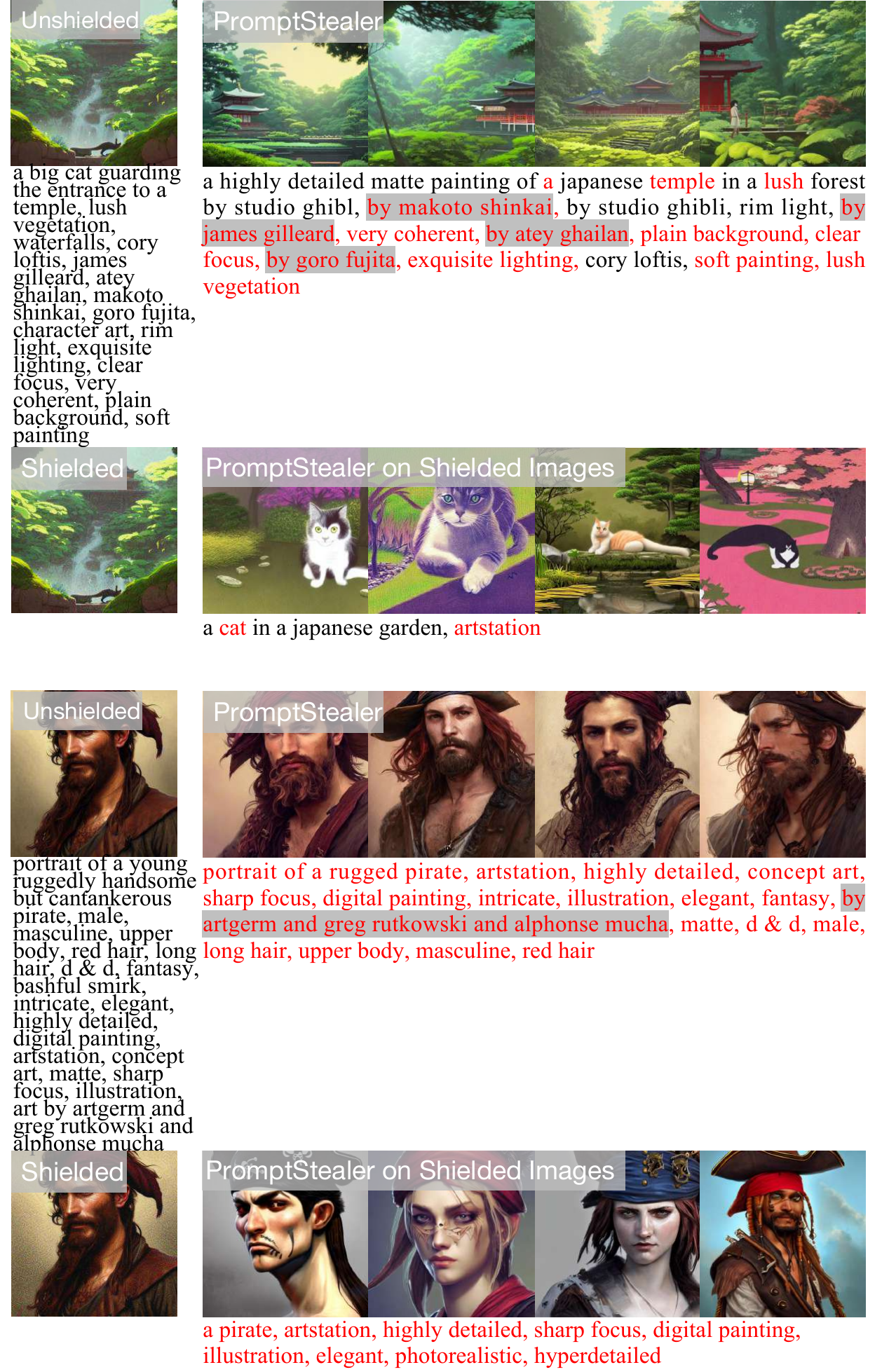}
\caption{Two defense examples of \defense (C\&W) against \attack.
The red color marks the correctly predicted modifiers. 
We highlight in grey the target artist modifiers stolen by \attack.}
\label{figure:defense_examples_mld_cw}
\end{figure}

C\&W is another representative method for generating adversarial examples.
Different from I-FGSM, C\&W utilizes two losses to control the attack effectiveness and utility.
In our experiments, we set the iterative step to 100, the learning rate to 0.05, and the loss trade-off hyperparameter to 0.001.
\autoref{table:defense_cw} reports the effectiveness of \defense (C\&W).
We observe that semantic, modifier, and image similarities decrease from 0.70, 0.43, and 0.80 to 0.58, 0.21, and 0.74, respectively.
Similar to the results shown in \autoref{table:defense_for_promptstealer}, the artist modifiers drop the greatest in similarity (from 0.49 to 0.13).
Regarding utility, we find C\&W also produces invisible perturbation.
The average MSE between the target and shielded images is 0.08.
\autoref{figure:defense_examples_mld_cw} shows two defense examples of \defense (C\&W) against \attack.

\begin{table}[!t]
\centering
\caption{The performance of \defense (C\&W) against \attack.
The second row refers to modifier categories.}
\label{table:defense_cw}
\tabcolsep 3.5pt
\scalebox{0.85}{
\begin{tabular}{l|ccccc}
\toprule
& \textbf{Semantic} & \textbf{Modifier} & \textbf{Image} & \textbf{Pixel} & \multicolumn{1}{|c}{\textbf{Human}}\\
\midrule
Unshielded & 0.70 & 0.43 & 0.80 & 0.90 & \multicolumn{1}{|c}{4.45}\\
Shielded & 0.58 & 0.21 & 0.74 & 0.88 & \multicolumn{1}{|c}{2.50}\\
\midrule
& \textbf{Artist} & \textbf{Medium} & \textbf{Flavor} & \textbf{Movement} & \textbf{Trending}\\
\midrule
Unshielded & 0.49 & 0.44 & 0.43 & 0.03 & 0.42\\
Shielded & 0.13 & 0.19 & 0.17 & 0.01 & 0.42\\ 
\bottomrule
\end{tabular}
}
\end{table}

\begin{figure}[!t]
\centering
\includegraphics[width=\linewidth]{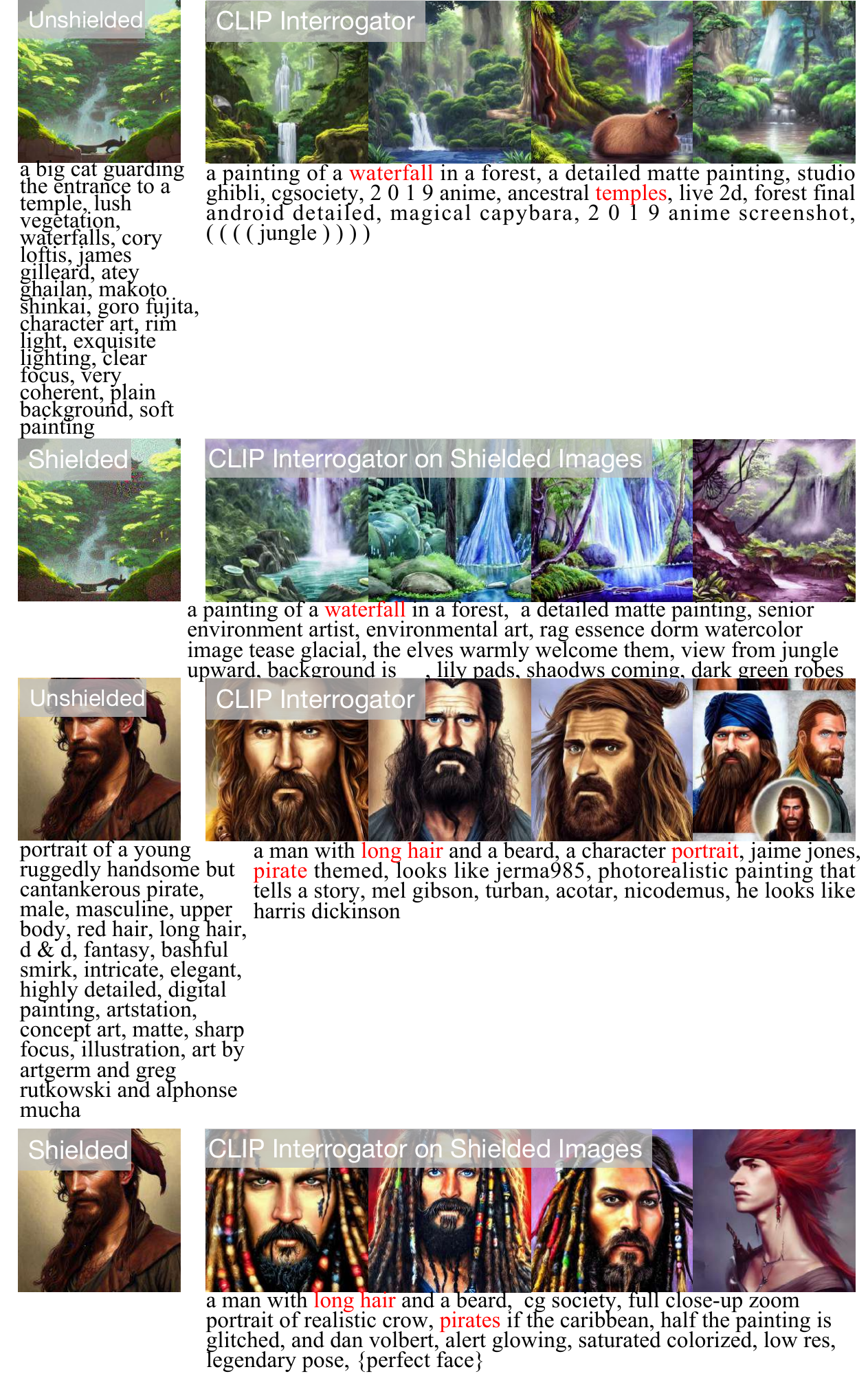}
\caption{Two defense examples of \defense against CLIP Interrogator.
The red color marks the correctly predicted modifiers.
We highlight in grey the target artist modifiers stolen by \attack.}
\label{figure:defense_examples_clip_interrogator}
\end{figure}

\section{\defense Against CLIP Interrogator}
\label{section:defense_clip_interrogator}

As \defense adds perturbations to mislead the attack model, it can be easily migrated to any other prompt stealing attacks relying on machine learning models.
In this experiment, we show the generality of \defense via applying it against CLIP Interrogator.
Specifically, we use I-FGSM as the optimization method as it produces smaller perturbations on the target images (see \autoref{section:defense_promptstealer_quantitative} and \autoref{section:defense_cw}).
We follow the same experiment settings in \autoref{section:defense_exp_settings}.

\mypara{Quantitative Evaluation}
Similar to \defense on \attack, we observe the semantic, modifier, and image similarities decrease significantly for CLIP Interrogator (see \autoref{table:defense_for_clip_interrogator}).
Specifically, the values of these metrics decrease from 0.52, 0.0108, and 0.77 to 0.33, 0.0059, and 0.51, respectively.
By examining the changes in different modifier categories, we find that the artist modifiers have the greatest drops in similarity (from 0.04 to 0.01).
Besides, the perturbation is minimal, i.e., the average MSE between the target and shielded images is 0.0002. 

\mypara{Qualitative Evaluation}
\autoref{figure:defense_examples_clip_interrogator} shows two defense examples of \defense against CLIP Interrogator.
In both examples, we find that the addition of imperceptible perturbation successfully misleads the attack model, i.e., the stolen images from the shielded images are quite different from the target images.
We also observe that the shielded images lead CLIP Interrogator to predict incorrect modifiers.
Take the first case in \autoref{figure:defense_examples_clip_interrogator} as an example.
The stolen prompt deriving from the shielded image contains modifiers like ``holography,'' ``m 1 abrams tank,'' ``incredible post - processing lighting,'' and ``brain scan.''
However, none of the modifiers belong to the target modifier set, resulting in the dissimilarity between the stolen images and the target image.

\begin{figure}[!t]
\centering
\begin{subfigure}{0.48\linewidth}
\includegraphics[width=\linewidth]{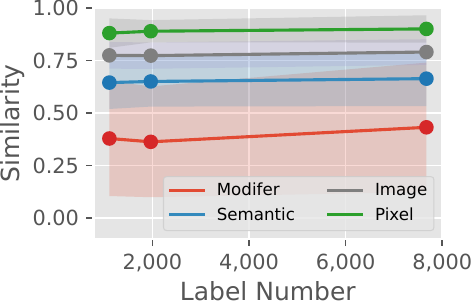}
\caption{Label number}
\label{figure:label_num}
\end{subfigure}
\begin{subfigure}{0.44\linewidth}
\centering
\includegraphics[width=\linewidth]{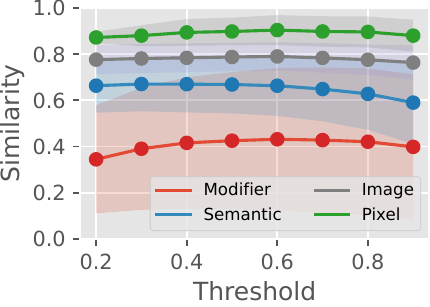}
\caption{Threshold}
\label{figure:threshold}
\end{subfigure}
\caption{Impacts of hyperparameters on \attack.}
\end{figure}

\begin{figure}[!t]
\centering
\includegraphics[width=\linewidth]{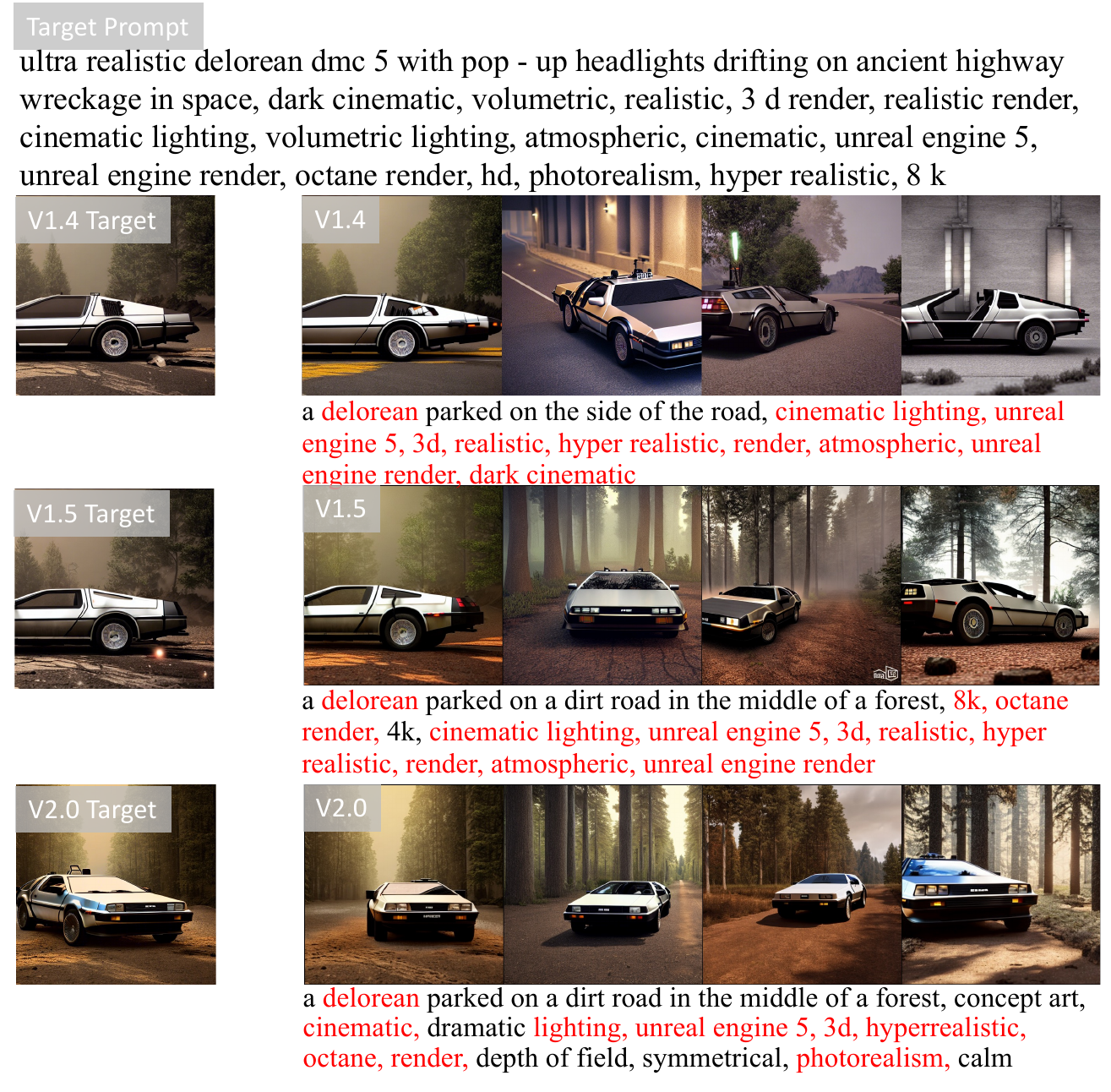}
\caption{One example of \attack across Stable Diffusion versions.}
\label{figure:sd_version}
\end{figure}

\begin{table}[!t]
\centering
\caption{The performance of \defense against CLIP Interrogator.
The second row refers to modifier categories.}
\label{table:defense_for_clip_interrogator}
\tabcolsep 3.5pt
\scalebox{0.85}{
\begin{tabular}{l|ccccc}
\toprule
& \textbf{Semantic} & \textbf{Modifier} & \textbf{Image} & \textbf{Pixel} & \multicolumn{1}{|c}{\textbf{Human}}\\
\midrule
Unshielded & 0.52 & 0.01 & 0.77 & 0.89 & \multicolumn{1}{|c}{2.95}\\
Shielded & 0.33 & 0.01 & 0.51 & 0.87 & \multicolumn{1}{|c}{2.00}\\
\midrule
& \textbf{Artist} & \textbf{Medium} & \textbf{Flavor} & \textbf{Movement} & \textbf{Trending}\\
\midrule
Unshielded & 0.04 & 0.01 & 0.00 & 0.01 & 0.03\\
Shielded & 0.01 & 0.01 & 0.00 & 0.00 & 0.03\\
\bottomrule
\end{tabular}
}
\end{table}

\begin{table}[!t]
\centering
\caption{The performance of \defense against \attack.
Adaptive represents the adaptive attack's performance.
The second row refers to modifier categories.}
\label{table:defense_adaptive}
\tabcolsep 3.5pt
\scalebox{0.85}{
\begin{tabular}{l|ccccc}
\toprule
& \textbf{Semantic} & \textbf{Modifier} & \textbf{Image}  & \textbf{Pixel} & \multicolumn{1}{|c}{\textbf{Human}}\\
\midrule
Unshielded & 0.70 & 0.43 & 0.80 & 0.90 & \multicolumn{1}{|c}{4.45}\\
Shielded & 0.62 & 0.74 & 0.71 & 0.88 & \multicolumn{1}{|c}{1.85}\\
Adaptive & 0.72 & 0.49 & 0.73 & 0.90 & \multicolumn{1}{|c}{4.05}\\
\midrule
& \textbf{Artist} & \textbf{Medium} & \textbf{Flavor} & \textbf{Movement} & \textbf{Trending}\\
\midrule
Unshielded & 0.49 & 0.44 & 0.43 & 0.03 & 0.42\\
Shielded & 0.06 & 0.53 & 0.79 & 0.06 & 0.48\\
Adaptive & 0.44 & 0.52 & 0.53 & 0.04 & 0.49\\
\bottomrule
\end{tabular}
}
\end{table}

\begin{table}[!t]
\centering
\caption{The criteria for scoring human-rated similarity.}
\label{table:human_rated_rules}
\scalebox{0.85}{
\begin{tabular}{p{.25\linewidth}|p{.7\linewidth}}
\toprule
\textbf{Level} & \textbf{Description} \\
\midrule
1 Not similar at all & The objects depicted are different, and the style is different.\\
2 Slightly similar & The objects depicted share some common elements or themes, but there are notable differences in both content and style.\\
3 Moderately similar & The objects depicted are similar in content or subject matter, but the styles are quite different.\\
4 Quite similar & The object depicted are similar in content or subject matter, but there are some differences in details, such as background color or minor elements.\\
5 Very similar & The objects depicted are almost identical in content, style, and details; they strongly resemble each other.\\ 
\bottomrule
\end{tabular}
}
\end{table}

\begin{figure*}[!t]
\centering
\includegraphics[width=0.9\linewidth]{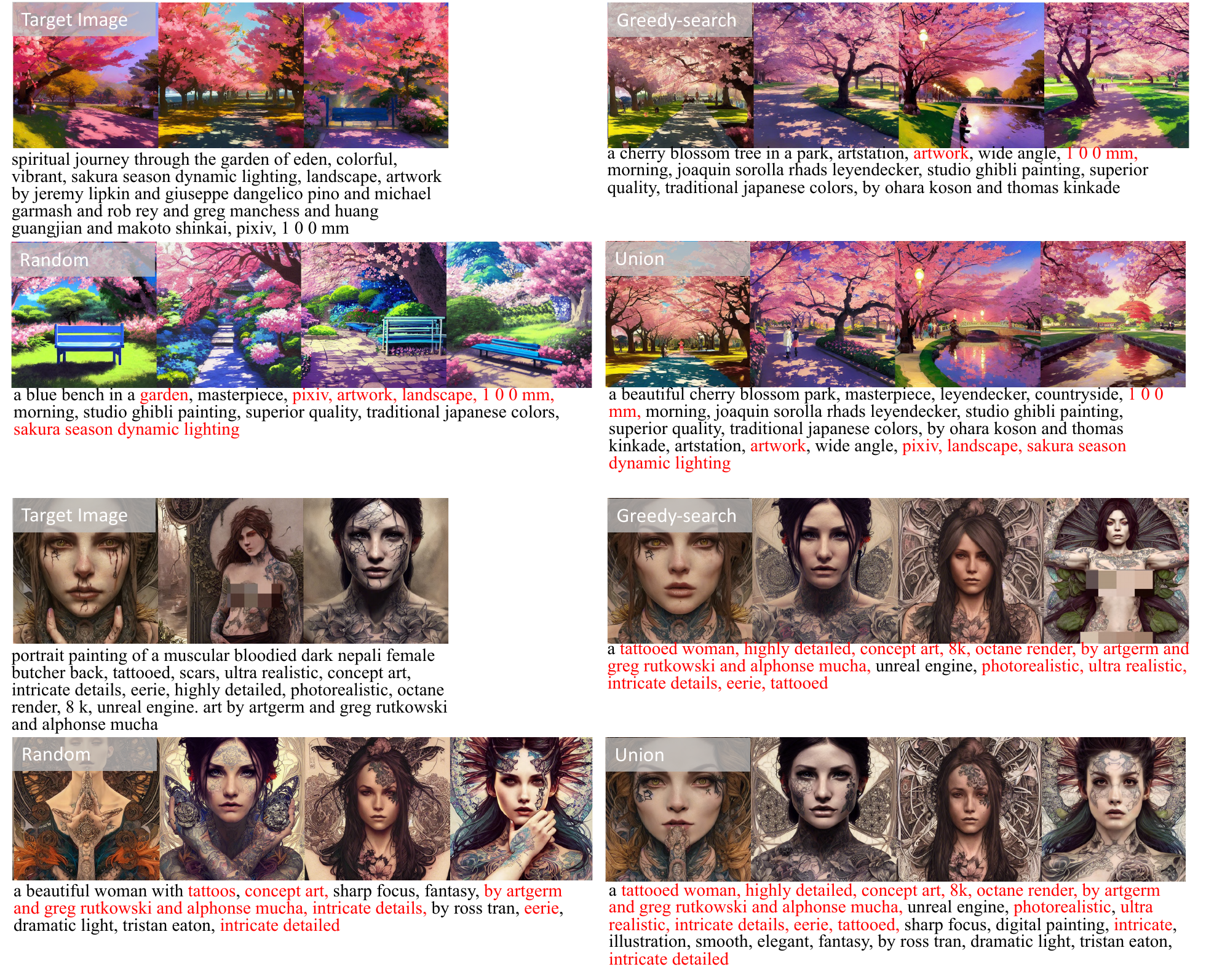}
\caption{Two attack examples of \attack on multiple target images.
We blurred the sexually explicit parts.}
\label{figure:multi_target}
\end{figure*}

\end{document}